\documentclass[conference]{IEEEtran}
\usepackage{graphicx}
\usepackage{amsmath}
\usepackage{lipsum}
\usepackage{amsthm}
\usepackage{color}
\usepackage{amsfonts}
\usepackage{caption}
\usepackage{blkarray}
\usepackage{mathtools}
\usepackage{siunitx}
\usepackage{amssymb}
\usepackage{subcaption}
\usepackage{placeins}
\usepackage{lettrine}
\usepackage{multicol, blindtext}
\usepackage{bbm}
\usepackage{cite}
\usepackage{algorithm}
\usepackage{mathrsfs} 
\usepackage{hyperref}
\usepackage{algpseudocode}
\usepackage{authblk}
\newtheorem{remark}{Remark}
\makeatletter

\newcommand{\Rmnum}[1]{\expandafter\@slowromancap\romannumeral #1@}
  
\newtheorem{theorem}{Theorem}
\newtheorem{corollary}{Corollary}
\newtheorem{proposition}{Proposition}
\newtheorem{assumption}{Assumption}
\newtheorem{definition}{Definition}
\newtheorem{lemma}{Lemma}

\makeatletter
\def\BState{\State\hskip-\ALG@thistlm}
\makeatother
\ifCLASSINFOpdf
\else
\fi

\begin{document}
%
\title{Age-Aware Stochastic Hybrid Systems: Stability, Solutions, and Applications}
\author[$\S$]{Ali Maatouk}
\author[*]{Mohamad Assaad}
\author[$\dagger$]{Anthony Ephremides}
\affil[$\S$]{Paris Research Center, Huawei Technologies, Boulogne-Billancourt, France}
\affil[*]{Laboratoire des Signaux et Syst\`emes, CentraleSup\'elec, Gif-sur-Yvette, France}
\affil[$\dagger$]{ECE Dept., University of Maryland, College Park, MD 20742}
\maketitle
\vspace{-25pt}
\begin{abstract}
In this paper, we analyze status update systems modeled through the Stochastic Hybrid Systems (\textbf{SHSs}) tool. Contrary to previous works, we allow the system's transition dynamics to be polynomial \color{black}functions of the Age of Information (\textbf{AoI}). This dependence allows us to encapsulate many applications and opens the door for more sophisticated systems to be studied. However, this same dependence on the AoI engenders technical and analytical difficulties that we address in this paper. Specifically, we first showcase several characteristics of the age processes modeled through the SHSs tool. Then, we provide a framework to establish the Lagrange stability and positive recurrence of these processes. Building on this, we provide an approach to compute the $m$-th moment of the age processes. Interestingly, this technique allows us to approximate the average age by solving a simple set of linear equations. Equipped with this approach, we also provide a sequential convex approximation method to optimize the average age by calibrating the parameters of the system. Finally, we consider an age-dependent CSMA environment where the back-off duration depends on the instantaneous age. By leveraging our analysis, we contrast its performance to the age-blind CSMA and showcase the age performance gain provided by the former.
\end{abstract}
\IEEEpeerreviewmaketitle
\section{Introduction}
In recent years, the proliferation of mobile devices, ubiquitous connectivity, and cheap hardware costs have paved the way for new real-time applications. Such applications include weather reporting, home appliance monitoring, vehicular networks, and many other up-and-coming applications. The common denominator for all these applications is their reliance on fresh data to achieve their optimal performance. In light of this, a metric of data freshness called the Age of Information (\textbf{AoI}) was proposed in \cite{6195689}. At any time $t$, if the freshest update
delivered to the destination had a timestamp of $U(t)$, then the
age at the destination side is
\begin{equation}
\Delta(t)=t-U(t).
\end{equation}
In other words, the AoI is the time elapsed since the moment that the freshest delivered update was generated. Accordingly, the AoI can be regarded as a measure of the information time-lag at the destination. Ever since its introduction, researchers have aimed to investigate this metric in various systems to gain insights on the means to achieve information freshness \cite{9380899,9241401,8406945,2018arXiv180103975J,Kadota:2018:SPM:3311528.3311547,popovski2021perspective,8945230,8756751,8613408,9155304,8006542,9137714,8764467
,6875100,8000687}. Notably, age-based metrics such as the average AoI \cite{6195689}, peak AoI \cite{6875100}, and non-decreasing age-functionals \cite{8000687} were minimized in an ample number of settings to achieve the desired freshness goal. Clearly, this type of analysis relies heavily on the tools that allow us to compute and analyze these age-based metrics. In other words, any progress in the tools mentioned above will enable the analysis of more sophisticated and elaborate systems. With this in mind, and given that the average AoI is the most commonly adopted age-based metric, we will focus on it in the remainder of this paper.


In a large part of the literature, the most fundamental tool that was leveraged to analyze the average age was the \emph{graphical decomposition method} \cite{6195689}. Given that the average age is nothing but the area below the instantaneous age curve, the method decomposes this area into several trapezoids. Afterward, the area of each trapezoid is written in function of key quantities such as packet inter-arrival time, queueing delay, transmission time, etc. Although this approach allows us to compute the average age in various settings, its usage can be limited  in systems where packets can be dropped (i.e., lossy systems) or
when a variety of events needs to be accounted for (e.g., a
random access environment). To address such limitations, another method, known as the Stochastic Hybrid Systems (\textbf{SHSs}) tool, has 
recently gained significant attention due to its versatility and simplicity. The use of the SHSs tool for AoI analysis was first done by Yates et al. in \cite{8469047} in the context of multi-source information flows system. In essence, an SHS revolves around two processes that interact with one another: 1) a continuous process, and 2) a discrete process, hence the name \emph{hybrid}. The discrete process captures events that could occur in the system, such as packet arrivals, successful packet transmissions, etc. On the other hand, the continuous process represents the age process of interest that evolves with time and is subject to potential changes at each transition of the discrete process. Under the constant transition rates assumption on the discrete process, Yates et al. have shown that the SHS model is stable and have provided an efficient way to calculate the average AoI. Accordingly, the work mentioned above has enabled the analysis of various systems using this tool. We cite for example Carrier-Sense-Multiple-Access (\textbf{CSMA}) environments \cite{9007478,9144109}, Non-Orthogonal-Multiple-Access environments \cite{8845254}, parallel servers networks \cite{8437907}, priority systems \cite{8437591,8849695}, Tandem queues \cite{8599728}, and many other systems.  Given the wide applicability of this tool, an interesting question arises: can we go beyond the constant transition rates assumption to more elaborate age-dependent transition rates? The motivation behind such a consideration is the vast amount of systems where the dynamics could change given the instantaneous AoI. For example, users with a higher age may be provided a higher access priority to the channel. Another example would be making the transmission time shorter for users with high age (e.g., by allocating a higher power to their transmission). Therefore, it is compelling to analyze such environments to open the door for more sophisticated systems to be studied. This is what this paper investigates and studies. To that end, the following are the key contributions of this paper:
\begin{itemize}
\item Starting with a general status update system modeled through the SHSs tool with age-dependent transition rates, we showcase the regularity of the stochastic processes involved. Additionally, we show that the SHS model satisfies the strong Markov property and several other critical characteristics. Equipped with these results, we formulate the differential equations that govern the evolution of the $m$-th moment of the age processes. 
\item The most fundamental step of our analysis is to showcase the stability of the differential equations involved. To tackle this, we distinguish in our analysis between two cases: 1) single discrete state, single age process systems, and 2) multiple discrete states, multiple age processes systems. This distinction is made to ease out the presentation of the paper as the analysis of the latter case is more complicated than the former. To that end, in the former case, we leverage Jensen's inequality to prove our desired stability results. In the latter case, and given its complexity, we focus on a particular age-aware CSMA environment and provide a framework to establish stability. It is worth noting that the proposed framework can be adapted to different systems by tweaking the proofs accordingly. In both cases, thanks to the stability results, we can assert the finiteness of the age moments for any arbitrary time instant $t$. 
\item Unfortunately, finiteness is not enough to prove the convergence of the differential equations in question. To ensure the convergence, we need to establish the ergodicity of the age processes on top of the Lagrange stability. With this goal in mind, we leverage the strong Markov property, Dynkin's formula and Fatou's lemma, along with Lyapunov functions to establish the positive recurrence of the age processes. Given the stability and positive recurrence, we can then assert the ergodicity of the age processes, affirming the convergence of the differential equations in the steady-state regime.  
\item Although the convergence of the differential equations is established, computing the moments of the age processes remains challenging due to the dependence of low-order moments on higher-order ones. To address this issue, we propose a moment closure technique based on a scaling of the differential equations. Interestingly, we show that our approach allows us to approximate the average age of the system with high accuracy by solving a set of linear equations. Given this fact, we also present a Sequential Convex Approximation (\textbf{SCA}) method to optimize the average age by calibrating the system's parameters. The convergence of the SCA procedure to a stationary point is then proven over any compact set of the parameters space.
\item Lastly, we implement the age-aware CSMA environment and contrast its performance to the age-blind CSMA. Interestingly, we demonstrate that the former shows significant performance gains, highlighting further the utility of the age-aware SHSs analysis.
\end{itemize}
The rest of the paper is organized as follows: Section \ref{shsintroduction} introduces the SHSs tool in general and the age-aware SHS in particular. In Section \ref{analysisoftheshs}, we prove several key characteristics of the SHS in question and lay out the fundamentals for Lagrange stability and positive recurrence. In Section \ref{bassingle} and \ref{bassmultiple}, we examine the stability and positive recurrence for 1) single discrete state, single age process systems, and 2) multiple discrete, states multiple age processes systems, respectively. In Section \ref{shssolutions}, we propose a moment closure technique to compute the average age and provide the SCA approach to optimize the average age of the systems in question. Lastly, the age-aware CSMA algorithm is implemented numerically in Section \ref{numericalimplementations}, while Section \ref{conclusionss} concludes the paper.\\
The notations adopted in the paper are as follows. We
use boldface to denote matrices and vectors. In addition, we let \textbf{I}$_n$ denote the $n\times n$ identity matrix. 
Also, let $\mathbb{E}[\cdot]$ denote the expectation, and $||\cdot||$ the vector norm.


\section{Introduction to SHSs}
\label{shsintroduction}
\subsection{General Settings}
Stochastic hybrid systems are dynamical systems that
combine 1) continuous change, 2) instantaneous change, and 3)
random effects. More precisely, the word ``hybrid" refers to these systems' ability to model the interaction between \textit{continuous} dynamics and \textit{discrete} dynamics. Some of the earliest references that study systems with these features include the work of Bellman \cite{10.1215/S0012-7094-54-02148-1} and Bergen \cite{1105035} in 1954 and 1960, respectively. Since then, the interest in this type of system skyrocketed due to the vast applications that fall under its umbrella. For example, SHSs were found to be useful in studying financial markets, air traffic management, communication networks, and even biological systems \cite{TEEL20142435}. Recently, and as has been pointed out in the introduction, the SHS tool was also found pivotal to model and analyze the AoI in status updates systems \cite{8469047}. 
 \color{black}

To understand these systems, let us define a jump process $q: [0,\infty)\mapsto\mathbb{Q}$ that we will refer to as the \emph{discrete} state. $\mathbb{Q}$ is a (typically finite) discrete set that amasses all the possible values of $q(t)$. Next, we let $\boldsymbol{x}: [0,\infty)\mapsto\mathbb{R}^n$ denote a stochastic process with piecewise continuous sample paths that we will call the \emph{continuous} state. An SHS is defined by a Stochastic Differential Equation (\textbf{SDE})
\begin{align}
\dot{\boldsymbol{x}}=f(q,\boldsymbol{x},t)+g(q,\boldsymbol{x},t)\dot{\boldsymbol{n}}, \quad &f:\mathbb{Q}\times\mathbb{R}^n\times[0,\infty)\mapsto\mathbb{R}^n,\nonumber\\& g:\mathbb{Q}\times\mathbb{R}^n\times[0,\infty)\mapsto\mathbb{R}^{n\times k},
\label{SDEeq}
\end{align}
a family of $L$ discrete transition/reset maps
\begin{equation}
(q,\boldsymbol{x})=\boldsymbol{\phi}_l(q^-,\boldsymbol{x}^-,t),  \quad \boldsymbol{\phi}_l:\mathbb{Q}\times\mathbb{R}^n\times[0,\infty)\mapsto\mathbb{Q}\times\mathbb{R}^n,
\end{equation}
and a family of $L$ transition intensities/rates
\begin{equation}
\lambda_l(q,\boldsymbol{x},t),  \quad \lambda_l:\mathbb{Q}\times\mathbb{R}^n\times[0,\infty)\mapsto[0,\infty), \quad \text{for }\:\: l=1,\ldots,L,
\end{equation}
where $\boldsymbol{n}$ denotes a $k$-vector of independent Brownian motion
processes. In other words, each component $n_i$ for $i=1,\ldots,k$ has the following characteristics
\begin{itemize}
\item $n_i(0)=0$
\item $n_i$ has independent increments: for every $t>0$, the future increments $n_i(t+u)-n_i(t), u\geq0$, are independent of the past values $n_i(s), s\leq t$. 
\item $n_i$ has Gaussian increments: $n_i(t+u)-n_i(t)$ is normally distributed with mean $0$ and variance $u$
\item $n_i(t)$ is continuous in $t$
\end{itemize}
In essence, the continuous process $\boldsymbol{x}(t)$ evolves according to the SDE reported in (\ref{SDEeq}). When a transition $l$ takes place, the corresponding reset map $\boldsymbol{\phi}_l(q^-,\boldsymbol{x}^-,t)$ is applied and the system continues evolving according to (\ref{SDEeq}), but starting from the new system state $(q,\boldsymbol{x})=\boldsymbol{\phi}_l(q^-,\boldsymbol{x}^-,t)$. Note that $(q^-,\boldsymbol{x}^-)$ denote both the discrete state and continuous state of the system just before the transition takes place. \color{black}Finally, to understand the transition rates, let us consider an elementary interval $(t,t+dt]$ and suppose that the system is in state $(q,\boldsymbol{x})$. Then, the probability of transition $l$ to take place is simply $\lambda_l(q,\boldsymbol{x},t)dt$. As can be seen, the SHS is quite a versatile modeling tool that can be used to represent a large variety of systems. In the next section, we will adapt it to the AoI framework in status updates systems. \color{black}
\subsection{Polynomial Age-dependent SHSs}
\label{depiction}
Let us take the general SHS model reported in the previous section and cater it to the special structure of the AoI framework in status updates systems. First, in status updates settings, the continuous process of the SHS model $\boldsymbol{x}(t)\in[0,\infty)^n$ denote a collection of age-related processes. For example, the components of $\boldsymbol{x}(t)$ could describe the age at the monitor of a certain stream or the age of a specific packet at time $t$. \color{black}On the other hand, the discrete process $q(t)\in\mathbb{Q}$ will denote the status of the network/system at hand. For example, transitions between the different possible values of $q(t)$ can occur when an event in the network happens, such as a successful packet transmission or a particular packet arrival. Next, given that $\boldsymbol{x}(t)$ tracks the evolution of a collection of age-related processes, we can assert that the SDEs governing its evolution are characterized by 
\begin{align}
&f(q,\boldsymbol{x},t)=\boldsymbol{b}_q,\nonumber\\
&g(q,\boldsymbol{x},t)=0,
\label{equationsss}
\end{align}
and $\boldsymbol{b}_q\in\{0,1\}^n$ is a binary vector for all $q\in\mathbb{Q}$. \color{black}To understand the choice of the above functions, we note that the evolution of the AoI is deterministic in nature as it can either grow linearly with time or keep the same value. For instance, if we track the age of a packet in a particular queue, its age grows linearly when the queue is non-empty. In contrast, if the queue is empty (i.e., there are no packets), its age stays equal to zero. Hence, the derivative of each component of $\boldsymbol{x}(t)$ can be either equal to $1$ or $0$. On top of that, given the deterministic nature of this evolution, we have $g(q,\boldsymbol{x},t)=0$. Thus far, we have not imposed any particularities on the status updates system in question. In other words, we did not impose any restrictions on the packet model, channel model, or queuing discipline. The first restriction we consider concerns the transition rates of the ensued SHS model. Specifically, we focus in this paper on the class of finite polynomial transition rates such that $\lambda_l(q,\boldsymbol{x},t)\geq 0$ for all $\boldsymbol{x}\in[0,\infty)^n$. \color{black}More precisely, the transition rates have the following form
\begin{equation}
\lambda_l(q,\boldsymbol{x},t)=\sum_{\boldsymbol{m}\in\mathbb{M}}a^{(l)}_{\boldsymbol{m}}\boldsymbol{x}^{\boldsymbol{m}}(t), \quad l=1,\ldots,L,
\end{equation}
where $\boldsymbol{m}\in\mathbb{N}^n$, $\mathbb{M}$ is the set of order vectors $\boldsymbol{m}$, and $\boldsymbol{x}^{\boldsymbol{m}}(t)=x_1^{m_1}(t)\ldots x_n^{m_n}(t)$. In previous works on SHSs for AoI analysis \cite{8469047}, the transition rates $\lambda_l$ were considered to be constant for all $\boldsymbol{x}$. However, by letting $\lambda_l(q,\boldsymbol{x},t)$ be dependent on $\boldsymbol{x}$, we can encompass a larger variety of systems and applications. One simple example of such systems being letting streams with higher age have more priority in accessing resources. It is worth noting that although we only consider polynomial functions, we can approximate various non-polynomials functions up to a certain accuracy through Taylor series \cite{lang_2012}. Lastly, to fully characterize our system, we consider that $\boldsymbol{\phi}_l(q,\boldsymbol{x},t)$ can have one of the two following forms
\begin{equation}
\boldsymbol{\phi}_l(q,\boldsymbol{x},t)=(q',\boldsymbol{x}'=\boldsymbol{A}_l\boldsymbol{x}) \quad \textnormal{or}\quad \boldsymbol{\phi}_l(q,\boldsymbol{x},t)=(q',\boldsymbol{x}'=\boldsymbol{c}_l), 
\end{equation}
where $(q'\boldsymbol{x}')$ denote the new state of the system after transition $l$ takes place, $\boldsymbol{c}_l\geq0$, and $\boldsymbol{A}_l\in \{0,1\}^{n\times n}$ is a binary reset matrix such that the sum of each row is less or equal to $1$\color{black}. Accordingly, $\boldsymbol{A}_l$ will allow us to incorporate transitions that lead to interchange between the components (e.g., $x'_1=x_2$) upon a packet delivery or when transitions have no effect on $\boldsymbol{x}$ (i.e., $\boldsymbol{A}_l=$ \textbf{I}$_n$). Consequently, we have $\boldsymbol{x}'=\boldsymbol{A}_l\boldsymbol{x}\leq\boldsymbol{x}$. On the other hand, reset maps of the form $\boldsymbol{x}'=\boldsymbol{c}_l$ allow us to reset the age processes to a predefined value.

\begin{figure}[!ht]
\centering
\includegraphics[width=.22\linewidth]{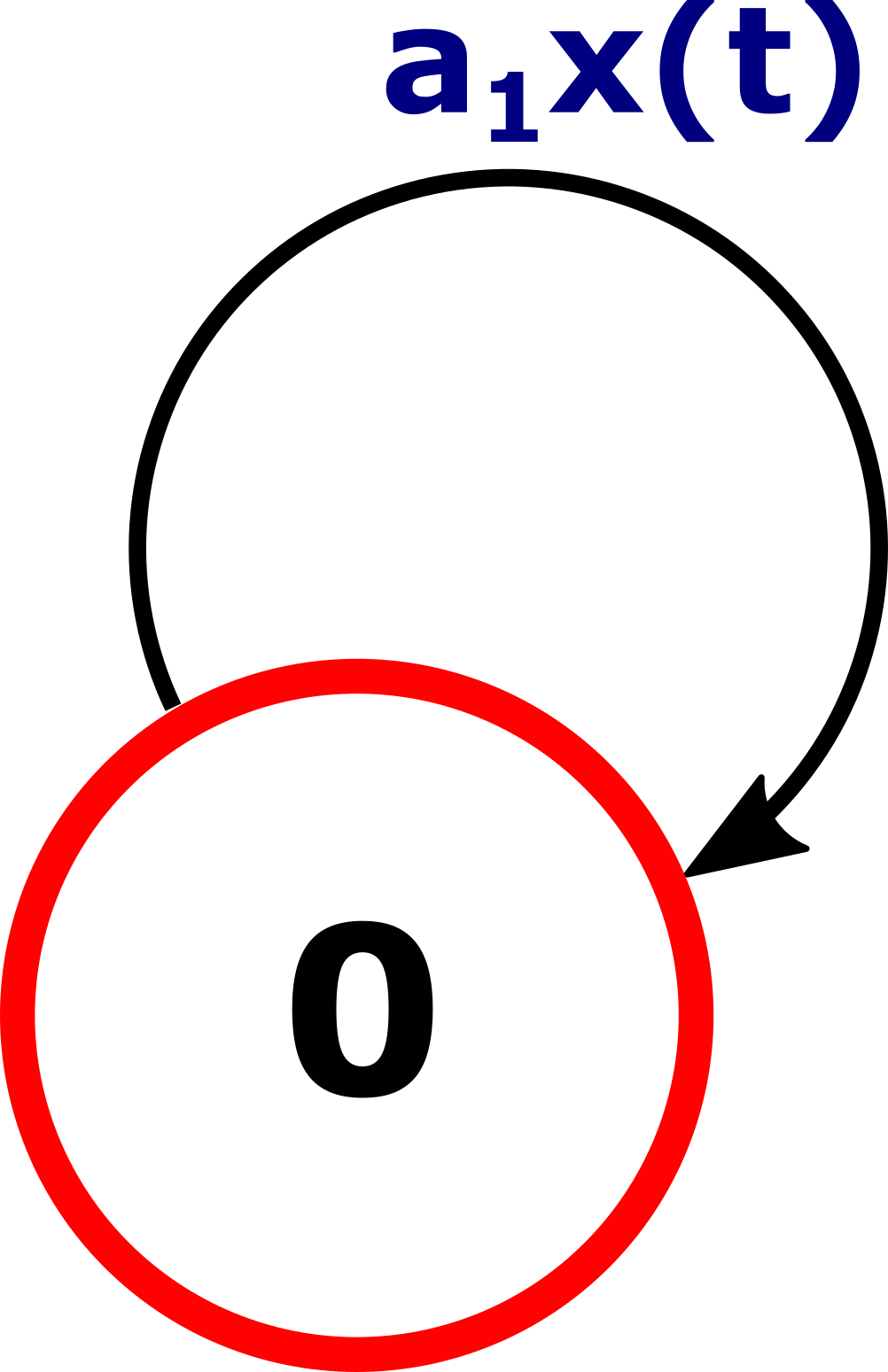}
\setlength{\belowcaptionskip}{-5pt}
\caption{Example of an age-dependent SHS.}
\label{illustrationexample}
\end{figure}

Although the case where $\lambda_l$ is independent of $\boldsymbol{x}(t)$ has been previously analyzed in \cite{8469047}, the above settings remain an open research area. The next section will showcase the need for a different methodology to study such systems compared to the independent case. Note that such an analysis will let us characterize/optimize the performance of an ample number of systems. We report below one of the simplest applications/examples of such systems. It is worth noting that the illustrative example (and later examples found in the paper) can be used as a reference for the readers to see how a status updates system can be fitted to an SHS model. \color{black}\\ 
\textbf{Illustrative Example.} Consider a scenario where $x(t)\in[0,\infty)$ denotes the AoI of a certain system at time $t$. To that end, we have
\begin{equation}
\dot{x}(t)=1, \quad \forall t\in[0,\infty).
\end{equation}
We consider that $q(t)$ can only have one value, such as $0$. In other words, $\mathbb{Q}$ is the singleton $\{0\}$. We suppose that as the age evolves, a self-transition counter starts ticking. Specifically, the transition intensity is equal to $\lambda_1(x(t),t)=a_1x(t)$ where $a_1>0$. When a transition occurs, the age $x(t)$ is reset to $0$. In other words, the reset function is defined as $\boldsymbol{\phi}_1(0,x,t)=(0,0)$. This simple scenario describes a transmitter-receiver zero delay system where, as the age grows, the transmitter decides with a higher probability to sample a specific physical process and sends a packet containing its value to the monitor side. An illustration of this example can be found in Fig. \ref{illustrationexample}. 
\\
Given the ability of age-dependent SHSs to model a wide variety of applications, a fundamental question arises: how can we analyze such systems? In the sequel, we answer this question and present a thorough analysis of the SHSs family in question. 
\section{Preliminaries to the SHSs Analysis}
\label{analysisoftheshs}
In status updates systems, our goal is to quantify an age-based metric for which the minimization allows us to achieve the freshness goal. Various metrics have been adopted throughout the literature, with the average AoI being the most common \cite{9380899}. To that end, given a certain status updates system modeled through the SHSs tool depicted above, our fundamental goal is to characterize the first-order statistics of the age processes (e.g., the average AoI). \color{black}Accordingly, let us consider the $n$-dimensional vector $\boldsymbol{m}=(m_1,m_2,\ldots,m_n)$ such that $m_i\in\mathbb{N}$ for $i=1,\ldots,n$ and $\sum_{i=1}^{n}m_i=m$. Moreover, we let $\boldsymbol{x}^{\boldsymbol{m}}(t)$ denote the monomial $x_1^{m_1}(t)x_2^{m_2}(t)\ldots x_n^{m_n}(t)$. With these quantities in mind, we define the moment of $\boldsymbol{x}(t)$ associated with $\boldsymbol{m}$ in state $\overline{q}\in\mathbb{Q}$ as

\begin{align}
\mu^{\boldsymbol{m}}_{\overline{q}}(t)&=\mathbb{E}[\boldsymbol{x}^{\boldsymbol{m}}(t)\delta_{\overline{q}q(t)}]=\mathbb{E}[x_1^{m_1}(t)x_2^{m_2}(t)\ldots x_n^{m_n}(t)\delta_{\overline{q}q(t)}]\nonumber\\&=\mathbb{E}[\boldsymbol{x}^{\boldsymbol{m}}(t)]\Pr(q(t)=\overline{q}), 
\label{mumstuff}
\end{align}
\color{black}where $\delta_{\overline{q}q(t)}$ is the Kronecker delta function. We also let $\boldsymbol{\mu}^{\boldsymbol{m}}(t)=[\mu^{\boldsymbol{m}}_{\overline{q}}(t)]_{\overline{q}\in\mathbb{Q}}$ be the vector that stacks the above moments. On top of that, we define the moment of $\boldsymbol{x}(t)$ associated with $\boldsymbol{m}$ as
\begin{equation}
\mu^{\boldsymbol{m}}(t)=\sum_{\overline{q}\in\mathbb{Q}}\mu^{\boldsymbol{m}}_{\overline{q}}(t).
\end{equation}
In other words, $\mu^{\boldsymbol{m}}(t)$ sums up all the contributions of the different discrete states in the evolution of the moment of $\boldsymbol{x}$ with respect to $\boldsymbol{m}$. With this in mind, if we are interested in the average AoI of the process $x_1$ for example, we set $\boldsymbol{m}^*$ to $(1,0,\ldots,0)$ and study the dynamics of the moment associated with $\boldsymbol{m}^*$. Note that $\mu^{\boldsymbol{m}}(t)$ depends on time and does not necessarily converge to a finite limit as $t\rightarrow \infty$. Therefore, to analyze and derive the average AoI (or any other moments of the age processes), a crucial step of the analysis consists of establishing the convergence of $\mu^{\boldsymbol{m}}(t)$ for any vector $\boldsymbol{m}$ \color{black}. We summarize below our approach to prove this
\begin{enumerate}
\item As a first step, we show that key properties hold for the SHS in question. These properties will allow us to establish an Ordinary Differential Equation (\textbf{ODE}) that governs the evolution of $\mu^{\boldsymbol{m}}_{\overline{q}}(t)$ for any $\boldsymbol{m}$ and $\overline{q}\in\mathbb{Q}$.
\item The next step of our analysis consists of proving the stability of the SHS in question. As will be explained later, there exist various forms of stability in the SHSs literature. However, we will be interested in establishing a specific form of stability: the Lagrange stability. Establishing the Lagrange stability ensures that the moments of the continuous process $\boldsymbol{x}$ are finite. The finiteness of the moments is a crucial technical step that precedes the derivation of the AoI moments\color{black}.
\item On top of the Lagrange stability, we also proceed with proving the positive recurrence of the SHS. Proving the positive recurrence is a fundamental step to conclude the ergodicity of the stochastic processes, thus allowing us to affirm the convergence of the moments of $\boldsymbol{x}$ in the steady-state regime. This convergence is also a critical prerequisite that precedes the derivation of the AoI moments as it assures us of their existence. \color{black}
\item Given the above results, we can assert the existence of a stationary distribution for the SHS. Accordingly, we can show the convergence of $\mu^{\boldsymbol{m}}(t)$ to a finite limit for any vector $\boldsymbol{m}$ as $t\rightarrow\infty$. 
\end{enumerate}
We start our analysis by providing the key properties verified by the SHS in question, establishing the ODEs, and laying the groundwork for the Lagrange stability and positive recurrence. From there onward, we proceed with our stability and positive recurrence analysis in two different cases 
\begin{enumerate}
\item \textbf{Single discrete, single age process}: This is the case where $|\mathbb{Q}|=1$ and $n=1$, where $|\cdot|$ denotes the cardinality of the set. The analysis provided in this scenario will pave the way for the latter more complicated case. 
\item \textbf{Multiple discrete states, multiple age processes}: In this more complicated case, we examine an age-dependent CSMA environment and showcase its Lagrange stability and positive recurrence. Although we focus on this particular example, the same analysis can be followed for different systems by adapting the proofs to the system of interest.
\end{enumerate}
With our work outlined, we can establish the key properties of the age-dependent SHS and lay the groundwork for our analysis. 
\subsection{Characteristics and Moment Dynamics}
The first key property of the system that we establish is its regularity. The regularity property assures us that the process $\{\boldsymbol{x}(t): t\geq0\}$ cannot blow up in finite time. 
\begin{definition}[Regularity \cite{yin2010}]
A stochastic process $\{\big(\boldsymbol{x}(t),q(t)\big): t\geq0\}$ is regular if and
only if for any $0<T<\infty$
\begin{equation}
\Pr\{\underset{0\leq t\leq T}{\sup}||\boldsymbol{x}^{\boldsymbol{x}(0),q(0)}(t)||=\infty\}=0,
\end{equation}
where $\boldsymbol{x}^{\boldsymbol{x}(0),q(0)}(t)$ denotes the value of $\boldsymbol{x}(t)$ given the initial conditions $(\boldsymbol{x}(0),q(0))$. 
\end{definition}
\begin{lemma}
The stochastic process $\{\big(\boldsymbol{x}(t),q(t)\big): t\geq0\}$ depicted in Section \ref{depiction} is regular.
\label{lemmaregular} 
\end{lemma}
\begin{IEEEproof}
First, we recall that $\boldsymbol{x}$ grows \emph{at most} linearly with time whenever there are no transitions. When a transition $l$ occurs, $\boldsymbol{x}(t)$ jumps to either a predefined vector $\boldsymbol{c}_l\geq 0$ or to $\boldsymbol{x}'=\boldsymbol{A}_l\boldsymbol{x}(t)$. Note that given the properties of $\boldsymbol{A}_l$ depicted in Section \ref{depiction}, we have $\boldsymbol{x}'=\boldsymbol{A}_l\boldsymbol{x}(t)\leq\boldsymbol{x}(t)$ for $l=1,\ldots,L$. With this in mind, we get that the modulus of $\boldsymbol{x}(t)$ verifies the following inequality
\begin{equation}
||\boldsymbol{x}(t)||\leq \max\{||\boldsymbol{x}(0)+\boldsymbol{1}T||,\underset{l}{\text{max}}||\boldsymbol{c}_{l}+\boldsymbol{1}T||\}, \quad t=0,\ldots,T.
\end{equation}
where $\boldsymbol{1}$ is a vector of dimension $n$ with all entries equal to $1$. Therefore, the process cannot blow up in finite time.
\end{IEEEproof}
Subsequently, we examine the strong Markov property of the process. This property will be of vital importance when we examine several types of stopping times in later sections of the paper.  
\begin{lemma}
The stochastic process $\{\big(\boldsymbol{x}(t),q(t)\big): t\geq0\}$, with natural filtration $\{\mathscr{F}_t\}_{t\geq0}$ verifies the strong Markov property. Specifically, for any stopping time $\tau$, conditioned on the event $\{\tau<\infty\}$ and given $\big(\boldsymbol{x}(\tau),q(\tau)\big)$, we have that $\big(\boldsymbol{x}(\tau+t),q(\tau+t)\big)$ is independent of $\mathscr{F}_{\tau}$ for each $t\geq0$.
\end{lemma}
\begin{IEEEproof}
To prove that our process verifies the strong Markov property, we note that the SHS reported in Section \ref{depiction} is a special case of the Piecewise Deterministic Markov Process (\textbf{PDMP}) introduced in the seminal paper of Davis in  \cite{10.2307/2345677}. Specifically, the process $\boldsymbol{x}(t)$ obeys to a deterministic ODE, and discrete transitions happen at random instants that induce changes to $\boldsymbol{x}(t)$. This coincides with the behavior of a PDMP (e.g., see \cite[Section~3.1]{refId0}). Now, given the regularity of the process shown in Lemma \ref{lemmaregular}, and by noting that the transition rates $\lambda_l(q,\boldsymbol{x},t)$ are continuous functions of $\boldsymbol{x}$, we can conclude that there are only finitely many jumps in finite time intervals. Therefore, Assumption 3.1 of \cite{10.2307/2345677} is verified. With this in mind, we can leverage the results of \cite[Section~4]{10.2307/2345677} to conclude that the stochastic process $\{\big(\boldsymbol{x}(t),q(t)\big): t\geq0\}$ verifies the strong Markov property. 
\end{IEEEproof}
The next step of our analysis consists of finding the dynamics that govern the evolution of $\boldsymbol{\mu}^{\boldsymbol{m}}(t)$. To do so, we first present several key results derived in \cite{HESPANHA20051353} that will allow us to proceed. 
\begin{assumption}[\cite{HESPANHA20051353}] The vector field $f(\cdot,\cdot,\cdot)$ is regular and the transition intensities $\lambda_l(q,\boldsymbol{x},t): \mathbb{Q}\times [0,\infty)^n \times [0,\infty)\mapsto [0,\infty)$ for $l=1,\ldots,L$ are measurable functions.
\label{assumptionhespana1}
\end{assumption}
\begin{assumption}[\cite{HESPANHA20051353}]
Let $\boldsymbol{\phi}^{\boldsymbol{x}}_l: \mathbb{Q}\times [0,\infty)^n \times [0,\infty)\mapsto [0,\infty)$ for $l=1,\ldots,L$ denote the projection of $\boldsymbol{\phi}_l$ into $[0,\infty)^n$. There exists a continuous function $\gamma_f:[0,\infty)\mapsto [0,\infty)$ and constants $c_f$, $c_{\phi}$ such that
\begin{equation}
||f(q,\boldsymbol{x},t)||\leq\max\{\gamma_f(t)||\boldsymbol{x}||,c_f\}, \quad||\boldsymbol{\phi}^{\boldsymbol{x}}_l(q,\boldsymbol{x},t)||\leq \max\{||\boldsymbol{x}||,c_{\phi}\},
\end{equation}
$\forall q\in\mathbb{Q},\boldsymbol{x}\in\mathbb{R}^{n},t\geq0,l\in\{1,\ldots,L\}.$
\label{assumptionhespana2}
\end{assumption}
\begin{theorem}
Suppose that Assumptions \ref{assumptionhespana1} and \ref{assumptionhespana2} hold. Let us define a test function $\psi(q,\boldsymbol{x},t):\:\mathbb{Q}\times [0,\infty)^n \times [0,\infty)\mapsto [0,\infty)$ as a continuously differentiable function with respect to its second and third arguments. We have that
\begin{equation}
\frac{d\mathbb{E}[\psi(q(t),\boldsymbol{x}(t),t)]}{dt}=\mathbb{E}[(L\psi)(q(t),\boldsymbol{x}(t),t)],
\end{equation}
where the operator $L$ is the SHS generator that maps for every function $\psi(q,\boldsymbol{x},t)$ the value $(L\psi)(q,\boldsymbol{x},t)$ as follows
\begin{align}
&\psi(q,\boldsymbol{x},t)\mapsto (L\psi)(q,\boldsymbol{x},t)= \frac{\partial \psi(q,\boldsymbol{x},t)}{\partial \boldsymbol{x}}f(q,\boldsymbol{x},t)+\frac{\partial \psi(q,\boldsymbol{x},t)}{\partial t}\nonumber\\&+\frac{1}{2}\text{trace}\big(\frac{\partial^2\psi(q,\boldsymbol{x},t)}{\partial\boldsymbol{x}^2}g(q,\boldsymbol{x},t)g(q,\boldsymbol{x},t)'\big)\nonumber\\&+\sum_{l=1}^{L}\lambda_l(q,\boldsymbol{x},t)[\psi(\boldsymbol{\phi}_l(\lambda_l(q,\boldsymbol{x},t),t)-\psi(q,\boldsymbol{x},t)].
\label{shsextendedgenerator}
\end{align}
\label{theoremtaba3hespana}
\end{theorem}
As one can see, the above theorem will allow us to evaluate the dynamics of the moments $\boldsymbol{\mu}^{\boldsymbol{m}}(t)$ by appropriately choosing the test functions $\psi$. However, to leverage this theorem, we first need to verify that our system verifies the key assumptions depicted above. 
\begin{lemma}
The stochastic process $\{\big(\boldsymbol{x}(t),q(t)\big): t\geq0\}$ depicted in Section \ref{depiction} verifies Assumptions \ref{assumptionhespana1} and \ref{assumptionhespana2}.  
\label{assumptionhespana}
\end{lemma}
\begin{IEEEproof}
As a first step, we note that the vector field $f(q,\boldsymbol{x},t)$ in eq. (\ref{equationsss}) is equal to $\boldsymbol{b}_q$, and therefore, the regularity assumption on the vector field $f(q,\boldsymbol{x},t)$ is trivial. Next, we recall that for the SHS depicted in Section \ref{depiction}, the transition intensities $\lambda_l(q,\boldsymbol{x},t)$ are polynomials in $\boldsymbol{x}$, which are continuous functions. With that in mind, and given that $\boldsymbol{x}\in[0,\infty)^n$, we can assert that these functions are measurable (see \cite[Proposition~2.3.1]{Cunha13}). Next, let us focus on bounding $\boldsymbol{\phi}^{\boldsymbol{x}}_l: \mathbb{Q}\times [0,\infty)^n \times [0,\infty)\mapsto [0,\infty)$ for $l=1,\ldots,L$. Note that if $\boldsymbol{\phi}_l(q,\boldsymbol{x},t)=(q',\boldsymbol{x}')$, then $\boldsymbol{\phi}^{\boldsymbol{x}}_l(q,\boldsymbol{x},t)=\boldsymbol{x}'$. Let also $\phi^{q}_l: \mathbb{Q}\times [0,\infty)^n \times [0,\infty)\mapsto \mathbb{Q}$ for $l=1,\ldots,L$ denote the projection of $\boldsymbol{\phi}_l$ into $\mathbb{Q}$. Given our system's dynamics, and as explained in the proof of Lemma \ref{lemmaregular}, we have
\begin{equation}
||\boldsymbol{\phi}^{\boldsymbol{x}}_l(q,\boldsymbol{x},t)||\leq \max\{||\boldsymbol{x}||,c_{\text{max}}\},\quad l=1,\ldots,L,
\end{equation}
where $c_{\text{max}}=\underset{l}{\text{max}}||\boldsymbol{c}_{l}||$. On top of that, we note that $f(q,\boldsymbol{x},t)=\boldsymbol{b}_q\leq \boldsymbol{1}$ for all $q\in\mathbb{Q},\boldsymbol{x}\in[0,\infty)^n,t\in[0,\infty)$. Therefore, our system satisfies the aforementioned assumptions.
\end{IEEEproof}
Given that our system verifies the key assumptions, we can now leverage Theorem \ref{theoremtaba3hespana} to find the dynamics governing the evolution of $\boldsymbol{\mu}^{\boldsymbol{m}}(t)$.
\begin{theorem} For any $\overline{q}\in\mathbb{Q}$, let $\psi_{\overline{q}}^{\boldsymbol{m}}(q(t),\boldsymbol{x},t)=\boldsymbol{x}^{\boldsymbol{m}}\delta_{\overline{q}q(t)}$. The SHS extended generator applied to this function is
\begin{align}
&L\psi_{\overline{q}}^{\boldsymbol{m}}(q(t),\boldsymbol{x},t)=\delta_{\overline{q}q(t)}\sum_{i=1}^{n}b^i_{\overline{q}}m_ix_i^{m_i-1}\prod_{j \ne i} x_j^{m_j}\nonumber\\&+\sum_{\hat{q}\in\mathbb{Q}}\delta_{\hat{q}q(t)}\big[\sum_{l\in\overline{\mathbb{L}}_{\overline{q},\hat{q}}}\lambda_l(\hat{q},\boldsymbol{x},t)(\boldsymbol{\phi}^{\boldsymbol{x}}_l(\hat{q},\boldsymbol{x},t))^{\boldsymbol{m}}\big]
\nonumber\\&-\delta_{\overline{q}q(t)}\sum_{l\in\mathbb{L}_{\overline{q}}}\lambda_l(\overline{q},\boldsymbol{x},t)\boldsymbol{x}^{\boldsymbol{m}},
\label{theoremlasese}
\end{align}
where $b^i_{\overline{q}}$ is the $i$-th component of $\boldsymbol{b}_{\overline{q}}$, $\mathbb{L}_{\overline{q}}$ is the set of transitions originating from $\overline{q}$, $\overline{\mathbb{L}}_{\overline{q},\hat{q}}$ is the set of transitions originating from $\hat{q}$ and ending in $\overline{q}$, and $(\boldsymbol{\phi}^{\boldsymbol{x}}_l(q(t),\boldsymbol{x},t)=\boldsymbol{x}')^{\boldsymbol{m}}=x'^{m_1}_1x'^{m_2}_2\ldots x'^{m_n}_n$. Moreover, the moment of $\boldsymbol{x}(t)$ associated with $\boldsymbol{m}$ in state $\overline{q}\in\mathbb{Q}$ verifies the following ODE
\begin{equation}
\frac{d\mathbb{E}[\mu^{\boldsymbol{m}}_{\overline{q}}(t)]}{dt}=\mathbb{E}[(L\psi_{\overline{q}}^{\boldsymbol{m}}(q(t),\boldsymbol{x},t)].
\label{derivativestufff}
\end{equation}
\label{theoremillustrative}
\end{theorem}
\begin{IEEEproof}
Given that our system verifies Assumptions \ref{assumptionhespana1} and \ref{assumptionhespana2}, we can leverage Theorem \ref{theoremtaba3hespana} to characterize the evolution of the moments dynamics of the system. Specifically, for any test function $\psi$, we have
\begin{equation}
\frac{d\mathbb{E}[\psi(q(t),\boldsymbol{x}(t),t)]}{dt}=\mathbb{E}[(L\psi)(q(t),\boldsymbol{x}(t),t)].
\end{equation}
To that end, we adopt the following family of test functions $\psi_{\overline{q}}^{\boldsymbol{m}}(q(t),\boldsymbol{x},t)=\boldsymbol{x}^{\boldsymbol{m}}\delta_{\overline{q}q(t)}=x_1^{m_1}x_2^{m_2}\ldots x_n^{m_n}\delta_{\overline{q}q(t)}$. As one can see in eq. (\ref{shsextendedgenerator}), the expression in the right hand side, which we will denote by $F(q,\boldsymbol{x})$, will depend on the value of the discrete state $q$. Given that our test function $\psi_{\overline{q}}^{\boldsymbol{m}}$ depends on $\overline{q}$, and given that our goal is to showcase the relationship between the different test functions $[\psi_{\overline{q}}^{\boldsymbol{m}}]_{\overline{q}\in\mathbb{Q}}$, we proceed similarly to \cite[Section VI]{1687493} and rewrite the SHS as follows
\begin{equation}
(L\psi)(q(t),\boldsymbol{x})=\sum_{\hat{q}\in\mathbb{Q}}\delta_{\hat{q}q(t)}F(\hat{q},\boldsymbol{x}).
\label{summingqhat}
\end{equation}
Let us now write $F(\hat{q},\boldsymbol{x})$ for the special case of $\psi_{\overline{q}}^{\boldsymbol{m}}(q(t),\boldsymbol{x},t)=\boldsymbol{x}^{\boldsymbol{m}}\delta_{\overline{q}q(t)}$. By leveraging eq. (\ref{shsextendedgenerator}), we obtain
\begin{align}
&F(\hat{q},\boldsymbol{x})=\delta_{\overline{q}\hat{q}}\sum_{i=1}^{n}b^i_{\overline{q}}m_ix_i^{m_i-1}\prod_{j \ne i} x_j^{m_j}\nonumber\\&+\sum_{l=1}^{L}\lambda_l(\hat{q},\boldsymbol{x},t)\Big[\delta_{\overline{q}\phi^q_l(\hat{q},\boldsymbol{x},t)}(\boldsymbol{\phi}^{\boldsymbol{x}}_l(\hat{q},\boldsymbol{x},t))^{\boldsymbol{m}}
-\delta_{\overline{q}\hat{q}}\boldsymbol{x}^{\boldsymbol{m}}\Big],
\label{Fdefinition}
\end{align}
where $(\boldsymbol{\phi}^{\boldsymbol{x}}_l(q(t),\boldsymbol{x},t)=\boldsymbol{x}')^{\boldsymbol{m}}=x'^{m_1}_1x'^{m_2}_2\ldots x'^{m_n}_n$ and $\phi^{q}_l$ is the projection of $\boldsymbol{\phi}_l$ into $\mathbb{Q}$. To obtain the results of the theorem, it suffices then to combine eq. (\ref{summingqhat}) and eq. (\ref{Fdefinition}) while keeping in mind the definition of the Kronecker delta function. 
Note that the differentiation of the test functions $\psi(\cdot,\cdot,\cdot)$ through the Kronecker functions will allow us to relate the different moments with one another as will be seen in later sections of the paper. 
\end{IEEEproof}
\color{black}

To understand the results of the above theorem, we apply it to the illustrative example depicted in Section \ref{depiction}. By doing so, we end up with the following expression for any $m\in\mathbb{N}^*$
\begin{equation}
\frac{d\mu^m(t)}{dt}=m\mu^{m-1}(t)-a_1\mu^{m+1}(t).
\label{odeillustrative}
\end{equation}
As can be seen from the equation above, the most challenging part about studying these systems is that the derivative of the moment of order $m$ depends on the moment of order $m+1$. By examining the results of Theorem \ref{theoremillustrative}, one can conclude that this is a consequence of the dependence of the transitions intensities on the process $\boldsymbol{x}(t)$. If we restrict ourselves to the case where the transition intensities $\lambda_l(q,\boldsymbol{x},t)$ are constants, the derivative of the moment of order $m$ will only depend on lower-order moments. In this case, to analyze the average AoI, for example, a linear system of a \emph{finite} number of equations can be formulated as done in the work of Yates et al. \cite{8469047}. Then, one can study the stability of this system through its eigenvalues. In contrast, in our case, the moment of order $1$ depends on the moment of order $2$, which itself depends on the moment of order $3$, etc. Accordingly, the same approach cannot be adopted, and we will have to leverage different tools to establish the stability of the SHS in question. To that end, we lay out the groundwork for our stability analysis in the following subsection.
\subsection{Lagrange Stability - Fundamentals}
In the literature of SHSs, stability has various definitions (Lagrange, Lyapunov, exponential stability, etc.). Depending on the framework, one may choose to adopt a definition of stability rather than the other. We refer the interested readers to the survey by Teel et al. in \cite{TEEL20142435}. In the sequel, we will be interested in the notion of Lagrange stability as it will elegantly allow us to proceed with our AoI analysis.  
\begin{definition}[Lagrange Stability]
For a stochastic hybrid system, the closed set $A\subset[0,\infty)^n$ is said to be Lagrange stable in the $m$-th mean $(m\geq1)$ if 
\begin{equation}
\underset{t\geq0}{\sup}\:\mathbb{E}[\underset{\boldsymbol{y}\in A}\inf ||\boldsymbol{x}(t)-\boldsymbol{y}||^{m}]<\infty.
\end{equation}
\end{definition}
It is common to study the stability for an equilibrium
point, often taken to be the origin. Without loss of generality, we adopt the same approach in our paper. In other words, we will be interested in proving that
\begin{equation}
\underset{t\geq0}{\sup}\:\mathbb{E}[||\boldsymbol{x}(t)||^{m}]<\infty.
\end{equation}
Concretely, the Lagrange stability consists of showing that the expected $m$-th power of the modulus of $\boldsymbol{x}(t)$ evolves over a bounded ball centered around the origin. Instead of examining the norm of $\boldsymbol{x}(t)$ to establish the stability, we focus on studying the moments of $\boldsymbol{x}(t)$ corresponding to a vector $\boldsymbol{m}\in\mathbb{N}^n$. As presented in the corollary below, proving the finiteness of these moments is sufficient to establish the Lagrange stability of the SHS. 
\begin{corollary}
If for any vector $\boldsymbol{m}=(m_1,m_2,\ldots,m_n)$ such that $m_i\in\mathbb{N}$ for $i=1,\ldots,n$ and $\sum_{i=1}^{n}m_i=m$ we have
\begin{equation}
\mu^{\boldsymbol{m}}(t)<\infty, \quad t\geq0,
\end{equation}
then the SHS is Lagrange stable.
\label{importantcorollary}
 \end{corollary}
 \begin{IEEEproof}
Our goal is to show that 
\begin{equation}
\underset{t\geq0}{\sup}\:\mathbb{E}[||\boldsymbol{x}(t)||^m]<\infty.
\end{equation}
To that end, we recall that given that the vector space to which $\boldsymbol{x}(t)$ belongs is finite dimensional, then all norms are equivalent. Accordingly, let us consider the $\infty$-norm $||\boldsymbol{x}(t)||_{\infty}=\underset{i}{\text{max}}\:x_i(t)$. Note that by assumption, we have that $\mu^{\boldsymbol{m}}(t)$ is bounded for $t\geq0$ and for any chosen vector $\boldsymbol{m}$. We can therefore consider the family of vectors $\boldsymbol{m}=m\boldsymbol{e_i}$, for $i=1,\ldots,n$ to conclude that
\begin{equation}
\mathbb{E}[x_i^m(t)]<\infty, \quad i=1,\ldots,n, \quad t\geq0.
\end{equation}
Given that this is true for any $i$, we can deduce that it is the case for $\mathbb{E}[\underset{i}{\text{max}}\:x_i^m(t)]$ for any time $t\geq0$. Therefore, we can conclude that
\begin{equation}
\underset{t\geq0}{\sup}\:\mathbb{E}[||\boldsymbol{x}(t)||^m_{\infty}]<\infty.
\end{equation}
\end{IEEEproof}
With the Lagrange stability defined, we can now lay out the foundation for another key aspect of our analysis: positive recurrence.
\subsection{Positive Recurrence - Fundamentals}
\label{postrecurrence}
This subsection is devoted to the definitions of recurrence and positive recurrence. To that end, let us consider a certain set $U=D\times J\subset [0,\infty)^n\times\mathbb{Q}$. We let
\begin{equation}
\sigma_U:=\inf\{t\geq0: (\boldsymbol{x}(t),q(t))\in U\},
\end{equation}
\begin{equation}
\tau_U:=\inf\{t\geq0: (\boldsymbol{x}(t),q(t))\notin U\},
\end{equation}
denote the entry and exit time out of the set $U$. Moreover, we let the recurrence time $\tau_{UU}$ denote the time for the stochastic $\{\big(\boldsymbol{x}(t),q(t)\big): t\geq0\}$ to return to $U$ given that it started in $U$. On top of that, we can define
\begin{equation}
\sigma_D:=\inf\{t\geq0: \boldsymbol{x}(t)\in D\},
\end{equation}
\begin{equation}
\tau_D:=\inf\{t\geq0: \boldsymbol{x}(t)\notin D\},
\end{equation}
as the entry and exit time out of $D$, regardless of the value of $q(t)$. The above quantities will be essential in our positive recurrence analysis in the sequel. With the above notions in mind, we provide the following definitions. 
\begin{definition}[Recurrence \cite{yin2010}]
For $U=D\times J\subset [0,\infty)^n\times\mathbb{Q}$, where $D$ is an open set with compact closure $\overline{D}$, let
\begin{equation}
\sigma^{\boldsymbol{x}(0),q(0)}_U:=\inf\{t\geq0: (\boldsymbol{x}(t),q(t))\in U\},
\end{equation}
 where the superscript $\boldsymbol{x}(0),q(0)$ denotes the the initial conditions. \color{black}A \textbf{regular} process $(\boldsymbol{x}(t),q(t))$ is said to be recurrent with respect to $U$ if $\Pr(\sigma^{\boldsymbol{x}(0),q(0)}_U<\infty)=1$ for any $(\boldsymbol{x}(0),q(0))\in D^c\times\mathbb{Q}$, where $D^c$ denotes the complement of $D$. Otherwise, the process is transient with respect to $U$.
\end{definition}
\begin{definition}[Positive Recurrence \cite{yin2010}]
A recurrent process with finite mean recurrence time for some set $U=D\times J\subset [0,\infty)^n\times\mathbb{Q}$, where $D$ is a bounded open set with compact closure, is said to be positive recurrent with respect to $U$. Otherwise, the process is null recurrent with respect to $U$. 
\end{definition}
In essence, the set $U$ is said to be recurrent if we are assured to eventually reach $U$ if we start outside of $U$. If, on top of that, the expected return time to $U$ is finite when starting from a point $(\boldsymbol{x},q)\in U$, we can conclude that the process is positive recurrent with respect to $U$. We are interested in positive recurrence since it is essential to establish the ergodicity of the process $(\boldsymbol{x}(t),q(t))$. The positive recurrence property allows us to ensure the ergodicity of $(\boldsymbol{x}(t),q(t))$, and therefore its steady-state convergence to a stationary measure. In fact, given the positive recurrence property, we can assure the existence of a unique stationary measure \cite[Theorem 4.3]{yin2010}. Additionally, we are guaranteed to converge to this aforementioned measure in the steady-state regime \cite[Theorem 4.4]{yin2010}. \color{black}
With these definitions laid out, we can now proceed with our analysis for the two cases of single and multiple discrete states as it has been previously outlined. 
\section{Single Discrete State, Single Age Process}
\label{bassingle}
In this section, we focus on the case where $|\mathbb{Q}|=1$ and $n=1$. Based on the general model depicted in Section \ref{depiction}, the transitions rates in this special case have the following form
\begin{equation}
\lambda_l(x(t),t)=\sum_{j=0}^{J_{l}}a^{(l)}_{j}x^{j}(t), \quad l=1,\ldots,L,
\end{equation}
where $J_l\geq0$ is the polynomial order of the rate of transition $l$, $\lambda_l(x(t),t)\geq0$ for $l=1,\ldots,L$, and $x(t)$ is the age process of interest that increases linearly with time (i.e., $f(x,t)=1$). Moreover, we consider the following transition reset functions 
\begin{equation}
\phi_l(x,t)=c, \quad l=1,\ldots,L,
\label{resetmapssingle}
\end{equation}
where $c\geq0$. There exists a range of applications for this SHS model, with the simplest being the illustrative example provided in Section \ref{depiction}. The first step of the subsequent analysis consists of applying Theorem \ref{theoremillustrative} on this specific family of systems. By doing so, and given that $n=1$, we end up with the following ODE
\begin{equation}
\frac{d\mu^m(t)}{dt}=m\mu^{m-1}(t)+\sum_{l=1}^{L}\sum_{j=0}^{J_{l}}a^{(l)}_jc^j\mu^{j}(t)-\sum_{l=1}^{L}\sum_{j=0}^{J_{l}}a^{(l)}_j\mu^{m+j}(t), 
\label{singlestatespecific}
\end{equation}
where $\mu^m(t)=\mathbb{E}[x^m(t)]$ is the age moment of order $m\geq1$.
\subsection{Lagrange Stability}
With the ODE established, we can now proceed with showcasing its Lagrange stability. To do so, we first study the particularity of the function $x^{y}$ for any $y\geq1$. To that end, we note that the function $g_{\text{power}}(x)=x^{y}$ is convex for $x\in[0,\infty)$ and for any $y\geq1$. With this in mind, we recall Jensen's ineqality for convex functions. Specifically, we have that for any real-valued stochastic variable $Y$ and convex function $g(\cdot)$
\begin{equation}
g(\mathbb{E}[Y])\leq \mathbb{E}[g(Y)].
\end{equation}
Given the convexity of the function $g_{\text{power}}(\cdot)$, let us apply Jensen's inequality on the stochastic process $x(t)$. We can conclude that for any $p\geq r\geq1$, we have
\begin{equation}
\mathbb{E}[x^r(t)]^{p/r}=[\mu^{r}(t)]^{p/r}\leq \mu^{p}(t)=\mathbb{E}[x^p(t)].
\label{illustrumomentcomparisonconvex}
\end{equation}
Given that the tricky part of the ODE reported in eq. (\ref{singlestatespecific}) is its dependence on higher-order moments, Jensen's inequality allows us to relate the different moments together to some extent. This is why Jensen's inequality was found to be essential in the analysis of various SHSs (we refer the reader to the survey in \cite{TEEL20142435}). With the above particularities of the SHS in question, we provide the following results on the finiteness of the age moments.
\begin{theorem} For any order $m\geq0$, the moments of the age process verify the following inequality
\begin{equation}
\mu^{m}(t)\leq U_m, \quad t\geq0,
\end{equation}
where $U_m$ is a finite positive number.
\label{stabilitysinglee}
\end{theorem}
\begin{IEEEproof}
The proof can be found in Appendix \ref{appendixstabilitysingle}.
\end{IEEEproof}
Given the above results, and by leveraging Corollary \ref{importantcorollary}, we can deduce the Lagrange stability of the system. 
\subsection{Positive Recurrence}
\label{positivesinglestate}
The stability results assure us that the moments are finite. However, the finiteness is, unfortunately, not enough to prove the convergence of the differential equation reported in eq. (\ref{singlestatespecific}) for any $m\geq1$. For example, the solutions to the ODE can oscillate within a certain interval without converging. To ensure the convergence, on top of the Lagrange stability, we need to establish the ergodicity of the stochastic process $\{x(t):t\geq0\}$. To do so, we will leverage several properties of the AoI evolution and the Lagrange stability results to prove the positive recurrence of the process. To proceed in this direction, let us first tweak the definitions reported in Section \ref{postrecurrence} to fit the single discrete state case. Let $D \subset [0,\infty)$ be a non-empty open interval with a compact closure. Let
\begin{equation}
\tau_D:=\inf\{t\geq0: x(t)\notin D\},
\end{equation}
\begin{equation}
\sigma_D:=\inf\{t\geq0: x(t)\in D\}.
\end{equation}
Concretely, the above two stopping times will allow us to define the notion of entry and exit time of a certain interval, as will be seen in the sequel. Given the evolution of the AoI, and the reset maps reported in (\ref{resetmapssingle}), it can be easily seen that the set $D_0=[0,c[$ is transient, where $c$ is the AoI value corresponding to the reset maps $\phi_l(x,t)$ for $l=1,\ldots,L$\color{black}. In fact, if $x(0)\in [c,\infty)$, then $\Pr(\sigma_{D_0}=\infty)=1$. Accordingly, we focus in the sequel on proving the positive recurrence of the process $x(t)$ with respect to $D\subset[c,\infty)$. To that end, we first show that the expected exit time of a non-empty interval $D$ with compact closure is finite. 
\begin{proposition}
 Let $D \subset [c,\infty)$ be a non-empty interval with compact closure $\overline{D}$. We have
 \begin{equation}
 \mathbb{E}_{x}[\tau_D]<\infty, \quad \forall x\in D, 
 \end{equation}
 where $\mathbb{E}_{x}$ is the expectation given the initial condition $x(0)=x$. 
 \label{exitfinitesingle}
\end{proposition}
\begin{IEEEproof}
The proof can be found in Appendix \ref{proofexitfinemnawalwahad}. 
\end{IEEEproof}
The above results allow us to assert that whatever the set $D$ with a compact closure we are in, we will eventually escape it in finite time. With the above results in mind, we can now prove the positive recurrence of the stochastic process $x(t)$ with respect to any non-empty open set $D \subset [c,\infty)$ with a compact closure $\overline{D}$. 
\begin{theorem}
The system in question is positive recurrent with respect to any non-empty open set $D\subset [c,\infty)$ with compact closure $\overline{D}$. In other words,
\begin{equation}
 \mathbb{E}_{x}[\tau_{DD}]<\infty, \quad \forall x\in D, 
 \end{equation}
 where $\tau_{DD}$ is the recurrence time of the set $D$.
 \label{positiverecurrencesinglestuff}
\end{theorem}
\begin{IEEEproof}
The proof can be found in Appendix \ref{proofpositiverecurrencesinglestuff}. 
\end{IEEEproof}
Given the positive recurrence of the process, along with the Lagrange stability that eliminates the possibility of infinite moments, we can be assured that as time passes by, we have
\begin{equation}
\boldsymbol{\mu}^{m}(t)\xrightarrow{t\rightarrow\infty}\boldsymbol{\mu}^{m}_{\infty},
\end{equation}
where $\boldsymbol{\mu}^{m}_{\infty}$ is the steady-state moment of order $m$. Given the ODE depicted in eq. (\ref{singlestatespecific}), we can stack the equations that the moments verify for any order $m\geq1$ and take their derivative to zero. Accordingly, we get that the vector $\boldsymbol{\mu}_{\infty}=[\mu^0(\infty),\mu^1(\infty),\ldots]$ verifies the following linear system
\begin{equation}
\boldsymbol{A}_{\infty}\boldsymbol{\mu}_{\infty}=\boldsymbol{b}_{\infty},
\end{equation}
where $\boldsymbol{A}_{\infty}$ and $\boldsymbol{b}_{\infty}$ are an infinite dimension matrix and vector respectively that incorporate the entries of the ODE for every $m\geq1$. Note that $\mu^0_{\infty}$ is trivially equal to $1$. Therefore, the remaining step of our analysis is to solve the above linear system and find an expression of the first-order moment of the age process of interest. This will be examined in Section \ref{shssolutions}.
\section{Multiple Discrete States, Multiple Age Processes}
\label{bassmultiple}
The previous section results were based on the use of Jensen's inequality thanks to the convexity of the function $g(x)=x^m$ for $x\in[0,\infty)$ and $m\geq1$. By considering the more general case where multiple age processes interact with one another, the same approach cannot be adopted as the function $g(\boldsymbol{x})=x_1^{m_1}\ldots x_n^{m_n}$ is not convex for $\boldsymbol{x}\in[0,\infty)^n$. Additionally, if $|\mathbb{Q}|>1$, one has to consider the contribution of each discrete state to the overall moment value. These two aspects greatly complicate the resulting ODEs and render the analysis of the system even more challenging. Given these difficulties, we focus in the rest of our analysis on a specific environment modeled through age-dependent SHS tools with multiple discrete states and age processes. The analysis done in the sequel can then be used as a roadmap by the reader to adapt it for their system of interest. \color{black}

\subsection{Environment Description}
\label{environmentdescrip}
One of the various applications of the SHS depicted in Section \ref{depiction} is a general age-aware CSMA environment. Specifically, let us consider a scenario where $n$ links (transmitter-receiver pairs) share a transmission medium. The transmitter side of each link sends status updates to its corresponding
monitor. However, due to interference, only one link can be
active at each time instant. Given that links typically exhibit random channel conditions, we assume that the transmission time of the packets of each link $i$ is exponentially distributed with an average of $\frac{1}{H_i}$. In CSMA environments, the transmitter senses the channel before attempting a transmission. If an interfering transmission is spotted, the transmitter waits for the channel to be free again. As for when the channel is found to be idle, the transmitter waits for a certain duration of time before transmitting, called the \emph{back-off} time. While waiting, it keeps sensing the environment to spot any conflicting transmission. If any interfering transmission is spotted, the transmitter immediately stops its back-off timer and waits for the medium to be free to resume it. In other words, the back-off timers of all links only tick when the channel is idle. After a successful transmission by a certain link, the transmitter side of this link generates a new back-off time to prepare for the next packet transmission. In other words, we assume links are always competing for the channel to send their packets. As for the packet arrivals, we suppose that the transmitter generates a new packet upon channel capture, which is then sent through the medium. Lastly, we consider that the back-off times are exponentially distributed with the back-off rate for link $i$ being $R_i(x_i(t))=a_ix_i(t)$, where $a_i>0$ is a fixed constant and $x_i$ for $i=1,\ldots,n$ denote the age at the monitor side of link $i$. In other words, the higher the age of a link, the more aggressive it is in its quest to capture the channel. 

Given that the back-off rate depends on the age and that the age evolves with time, practical implementation issues arise. In fact, although mathematically we can model an exponential back-off time with a time-variable rate, we need to consider how to implement such a mechanism in practice. To that end, we summarize how to achieve this in the following. \\
\textbf{Practical implementation}: To explain the proposed implementation, let us consider an exponential clock $X$ with rate $\lambda>0$. The probability that a tick takes place in the interval $(t,t+\Delta t]$ is
\begin{equation}
\Pr(X\leq t+\Delta t|X>t)=\frac{\Pr(t<X\leq t+\Delta t)}{\Pr(X>t)}=1-\exp{(-\lambda\Delta t)}.
\end{equation}
Therefore, if we consider an elementary time interval $\Delta t$ where the age can be considered constant for its duration (and accordingly the back-off rate is constant too), it suffices to let users access the channel during this elementary time-interval with a probability 
\begin{equation}
p_i(t)=1-\exp(-a_ix_i(t)\Delta t)
\end{equation}
to implement the age-dependent back-off rate environment. With that in mind, we recall that in every practical CSMA protocol, time is discretized, and the duration of each time slot is defined as $T_{\text{slot}}$. This slot duration is pre-determined based on wave propagation delay and various other factors. For example, in IEEE 802.11n, it is equal to $9\mu s$ \cite{5307322}. Accordingly, to implement our approach, we suppose that when the channel is free, transmitters access the channel at each time slot with a probability 
\begin{equation}
p_i(t)=1-\exp(-a_ix_i(t)T_s)
\end{equation}
using, for example, the Request To Send/Clear To Send (\textbf{RTS/CTS}) mechanism of the IEEE 802.11 protocol. Then, when a link captures the channel, the rest of the links stay silent, waiting for its transmission to finish. Given that $T_{\text{slot}}$ is typically small, we can assert that this approach allows us to implement the age-dependent rate approach practically. Note that for simplicity, we will be ignoring the possible effect of collisions in our analysis, and we refer the readers to \cite{9007478} where it was discussed how to incorporate this practical issue by imposing a simple upper bound constraint on the back-off rate. Equivalently, we can handle the collisions in this case by upper bounding the values of $a_i$ for $i=1,\ldots,n$, for example, in the framework provided in Section \ref{parametrssoptimization}. 

To the best of the authors' knowledge, this is the first work that theoretically investigates a CSMA environment where the back-off duration depends on the instantaneous age of the link. In previous works on CSMA environments, the average AoI was optimized by tweaking a constant parameter that does not depend on the instantaneous age (e.g., \cite{9007478}). The same goes for most of the works on random access environments, where the AoI was optimized by calibrating age-blind parameters (e.g., \cite{2018arXiv180306469T}). In some other works, a threshold approach was adopted to incorporate the instantaneous AoI in the analysis. For example, in \cite{2020arXiv200709197T}, a slotted ALOHA environment was studied where only users with an age larger than a specific threshold $\Gamma$ access the channel. Given that our work constitutes the first age-aware CSMA environment, a fundamental question arises: how much performance gain can be achieved by letting these back-off timers depend on the AoI? To answer this question, we will model our system and leverage our SHS results from Section \ref{analysisoftheshs} to examine the average AoI in this environment. This further highlights the importance of the age-dependent SHSs framework as it allows us to answer this type of fundamental question. 
\subsection{Support of the Age Processes}
\label{recurrenceeee}
\begin{figure}[!ht]
\centering
\includegraphics[width=.6\linewidth]{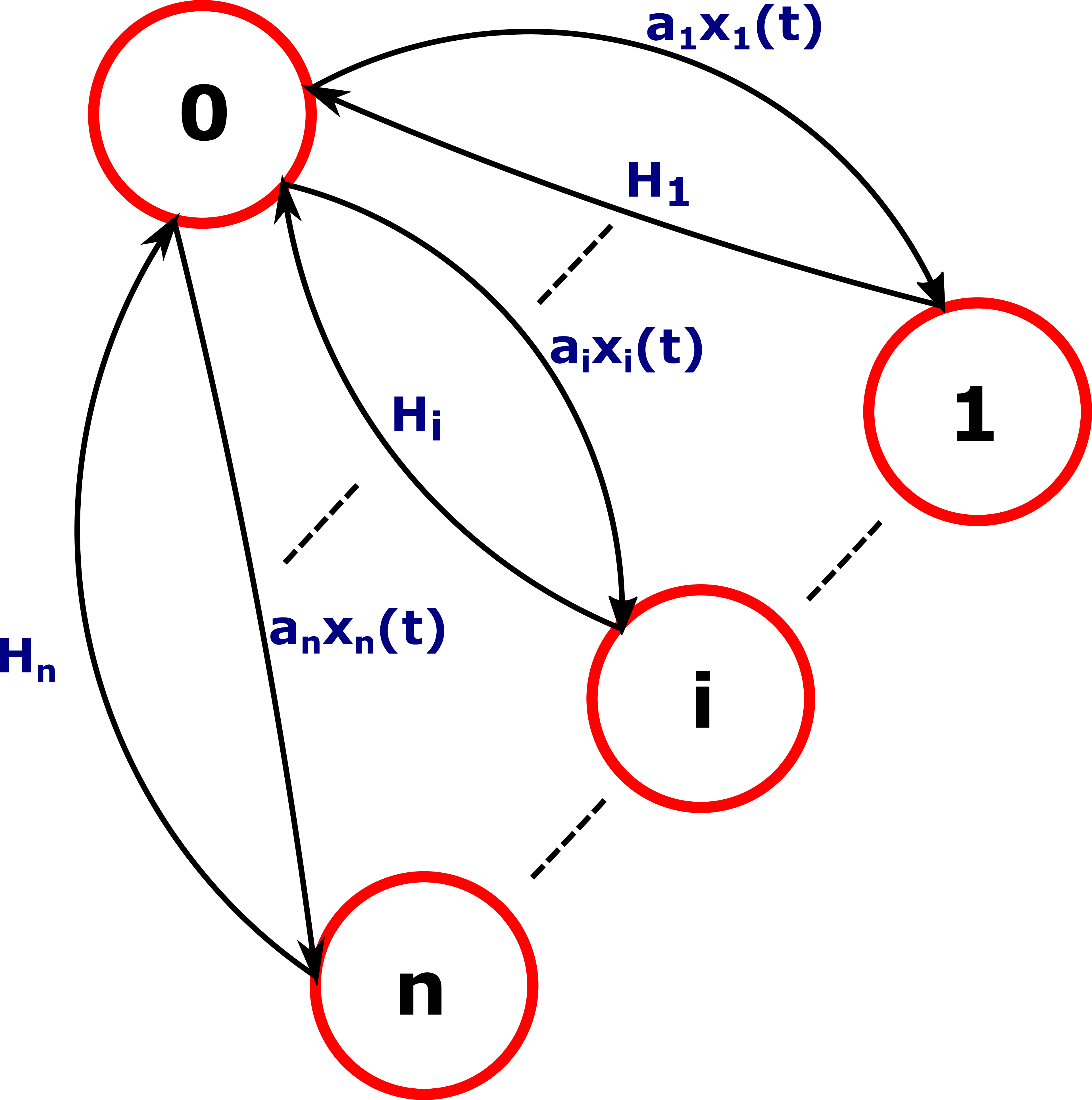}
\setlength{\belowcaptionskip}{-5pt}
\caption{Illustration of the age-aware CSMA.}
\label{generalcsmaxxx}
\end{figure}

To start our analysis, we model our system using the SHSs tool. To that end, we first note that our system falls under the umbrella of polynomial age-dependent SHSs models of Section \ref{depiction} as will be depicted in the following. First, let us consider the continuous process $\boldsymbol{x}(t)\in[0,\infty)^{2n}$ where $x_i(t)$ for $i=1,\ldots,n$ denotes the age at the monitor side of link $i$ at time $t$ and $x_{n+i}(t)$ for $i=1,\ldots,n$ denotes the age of the packet at the transmitter side of link $i$ at time $t$. On top of that, we recall from Section \ref{depiction} that $q(t)\in\mathbb{Q}$ is a discrete process that captures the status of the network in question. In our case, we set $q(t)$ to $0$ when the network is idle (i.e., when no link is transmitting). When link $i$ captures the medium and starts transmission, then $q(t)$ will be equal to $i$. Accordingly, we have $\mathbb{Q}=\{0,1,\ldots,n\}$. Given the dynamics of the system described in the previous section, an illustration of the possible transitions between the different values of $q(t)$ and their rates can be highlighted in Fig. \ref{generalcsmaxxx}. 
\\
With the continuous and discrete processes clarified, what remains is to characterize the other components of the SHS model. To that end, let us note that 
\begin{itemize}
\item The age at the monitor of all links grows linearly with time regardless of the value of the discrete process $q(t)$. On the other hand, the age of the packet at transmitter $k$ grows linearly with time only if a packet exists in its system. Therefore, $\boldsymbol{x}(t)$ evolves according to eq. (\ref{equationsss}) such that
\begin{align}
b^i_q&=1,\quad \textnormal{for } i=1,\ldots,n,n+q,\nonumber\\
b^i_q&=0,\quad \textnormal{otherwise},
\end{align}
where $b^i_q$ is the i-th component of the vector $\boldsymbol{b}_q$ for any $q\in\mathbb{Q}$. 
\item Transitions originating from the state $\{0\}$ take place when a link captures the channel. These transitions do not affect the value of the age processes (i.e., the corresponding transition matrix $\boldsymbol{A}_l\in\{0,1\}^{2n\times 2n}$ reported in Section \ref{depiction} is equal to \textbf{I}$_{2n}$). 
\item Finally, when a successful transmission happens, for example a transition from state $k\neq0$ to state $0$, the age of link $k$, denoted by $x_k$, becomes equal to $x_{k+n}$ and $x_{k+n}$ becomes equal to $0$. The matrices $\boldsymbol{A}_l$ corresponding to these transitions can be formulated based on this fact while noting that all the other age processes remain unchanged. 
\end{itemize}
Given all the above, we have a complete characterization of the SHS model of the age-dependent CSMA and all its elements that were reported in Section \ref{depiction}.
\color{black}Consequently, we can now determine the support of the stochastic process $\{\big(\boldsymbol{x}(t),q(t)\big): t\geq0\}$. In other words, the smallest closed set to which $(\boldsymbol{x}(t),q(t))$ belong with probability $1$. To that end, we first note that the age of the packet at transmitter $k$ is equal to $0$ when $q(t)\neq k$. Expressly,
\begin{equation}
x_{k+n}=0, \quad\textnormal{for } q(t)\neq k, k=1,\ldots,n.
\end{equation}
On top of that, given that the age of the packet at transmitter $k$ increases from $0$ solely when $q(t)$ transitions to state $k$, we can also conclude that for $q(t)=k$, we have that
\begin{equation}
x_{k+n}<x_i, \quad \forall i\in\{1,\ldots,n\}, \quad \textnormal{for } k=1,\ldots,n. 
\end{equation}
With this in mind, we can summarize below the set of states $D_{\textnormal{definition}}\times\mathbb{Q}$ that the process belong to
\begin{equation}
\begin{cases}
    [0,\infty)^n\times\{0\}\times\ldots,\{0\},& \textnormal{for } q(t)=0,  \\
   [0,\infty)^n\times [0,\infty)\times\{0\}\times\ldots,\{0\}, & \textnormal{for } q(t)=1,  \\
    \textnormal{subject to } x_{n+1}<x_i,\:\: \forall i\in\{1,\ldots,n\}. \\
    
   \vdots  &   \vdots\\
   [0,\infty)^{n}\times\{0\}\times\ldots,\{0\}\times [0,\infty), & \textnormal{for } q(t)=n,  \\  
   
    \textnormal{subject to } x_{2n}<x_i,\:\: \forall i\in\{1,\ldots,n\}. \\
  \end{cases}
\end{equation}
The above depiction of the support will be useful when we establish the positive recurrence of the stochastic process $\{\big(\boldsymbol{x}(t),q(t)\big): t\geq0\}$ in the next section.
\subsection{Positive Recurrence}
\label{multiplepositiverecurrence}
With the support of the stochastic process defined, we proceed with establishing its positive recurrence. Note that unlike the single state case studied in Section \ref{bassingle}, we start our analysis with the positive recurrence of the stochastic process instead of the Lagrange stability. This is a consequence of the fact that studying the ODEs that govern the moment dynamics in the multiple states case is rather challenging. Instead, we will establish the Lagrange stability using the positive recurrence results established in this section. To that end, the first step of our analysis is to show that the expected exit time of any non-empty set $D$ with a compact closure is finite. 
\begin{proposition}
 Let $D \subset [0,\infty)^{2n}$ be a non-empty set with compact closure $\overline{D}$. We have
 \begin{equation}
 \mathbb{E}_{\boldsymbol{x},q}[\tau_D]<\infty, \quad \forall(\boldsymbol{x},q)\in D\times\mathbb{Q}, 
 \end{equation}
 where $\mathbb{E}_{\boldsymbol{x},q}$ is the expectation given the initial conditions $(\boldsymbol{x},q)$. 
 \label{exittimemultiple}
\end{proposition}
\begin{IEEEproof}
The proof can be found in Appendix \ref{proofexittimemultiple}. 
\end{IEEEproof}
The above results allow us to assert that whatever the set $D$ with a compact closure we are in, we will eventually escape it in finite time. This will be crucial for the sequel. The next step in our preparation for our positive recurrence proof, we show below that we are guaranteed to have an age at the monitor of each link strictly larger than $0$. 
\begin{lemma}
For any initial state $(\boldsymbol{x},q)$, we have
\begin{equation}
{\Pr}_{\boldsymbol{x},q}(x_i(t)=0)=0,\quad \textnormal{for } i=1,\ldots,n,\quad t>0,
\end{equation}
where ${\Pr}_{\boldsymbol{x},q}$ denotes the probability given the initial condition $(\boldsymbol{x},q)$. \color{black}
\label{differentthan0}
\end{lemma}
\begin{IEEEproof}
To prove this, we note that the components $x_i(t)$ of the vector $\boldsymbol{x}(t)$ evolve in a deterministic way in each discrete state. In particular, they grow linearly with time. On top of that, after a transition from any state $k\neq0$ to state $0$, the age process $x_k(t)$ inherits the value $x_{n+k}(t)$. Note that at the point of transition, $x_{n+k}(t)$ will be equal to the time spent in state $k$. Given that the time spent in state $k$ is exponentially distributed with rate $H_k$, we have $\Pr(H_k=0)=0$. This concludes our proof. 
\end{IEEEproof}
Given the above lemma, and without loss of generality, we will assume that the initial state $\boldsymbol{x}(0)$ verifies $x_i\neq0$ for $i=1,\ldots,n$. Next, we establish the positive recurrence of the discrete component of $\{\big(\boldsymbol{x}(t),q(t)\big): t\geq0\}$. This will pave the way for us to prove the positive recurrence of the overall process.  
\begin{proposition}
The discrete stochastic process $\{q(t):t\geq0\}$ is positive recurrent with respect to any state $q\in\mathbb{Q}$. 
\label{qtisreccureent}
\end{proposition}
\begin{IEEEproof}
The proof can be found in Appendix \ref{proofqtisreccureent}. 
\end{IEEEproof}
With this in mind, we can combine all the propositions of this section to provide our main positive recurrence results below.
\begin{theorem}
For any set $U=D\times\{q\}\in D_{\textnormal{definition}}\times\mathbb{Q}$ where $D$ is a non-empty open set with compact closure $\overline{D}$, we have
\begin{equation}
\mathbb{E}_{\boldsymbol{x},q}[\tau_{UU}]<\infty
\end{equation}
for any initial state $\boldsymbol{x}\in D$. 
\label{ahamtheoremblpositive}
\end{theorem}
\begin{IEEEproof}
The proof can be found in Appendix \ref{proofahamtheoremblpositive}.
\end{IEEEproof}

With the positive recurrence of the stochastic process $\{\big(\boldsymbol{x}(t),q(t)\big): t\geq0\}$ being established, we can now focus on showing the Lagrange stability of the process.
\subsection{Lagrange Stability}
\label{lagrangeeee}
Our goal in this section is to establish the Lagrange stability. Proving this form of stability ensures us that the moments of $\boldsymbol{x}(t)$ with respect to any vector $\boldsymbol{m}$ are finite. In the previous case, this was done by leveraging Jensen's inequality. However, in this multi-process case, different machinery has to be leveraged to prove such results, as will be detailed in the following theorem.\color{black}
\begin{theorem} 
For any vector $\boldsymbol{m}$, the moment of $\boldsymbol{x}(t)$ associated with $\boldsymbol{m}$ verifies the following inequality
\begin{equation}
\mu^{\boldsymbol{m}}(t)=\mathbb{E}[\boldsymbol{x}^{\boldsymbol{m}}(t)]\leq U_{\boldsymbol{m}}, \quad t\geq0,
\end{equation}
where $U_{\boldsymbol{m}}$ is a finite positive number.
\label{stabilitymultipleee}
\end{theorem}
\begin{IEEEproof}
The proof can be found in Appendix \ref{ekherproof}.
\end{IEEEproof}
Given the above results, and by leveraging Corollary \ref{importantcorollary}, we can deduce the Lagrange stability of the system. With the positive recurrence and Lagrange stability results established, we can conclude the convergence of the moments in the steady-state regime. In the next section, we provide further details on the ODEs that govern the dynamics of these moments in order to establish the equations that the steady-state moments verify.

\subsection{Moment Dynamics}
As previously discussed, we can obtain the ODEs that the moments verify by applying Theorem \ref{theoremillustrative} on the system in question. To that end, let us apply Theorem \ref{theoremillustrative} to find the differential equations that the moments of $\boldsymbol{x}$ verify. To do so, let us define the following vectors $\overset{0}{\boldsymbol{m}}$ and $\overset{k}{\boldsymbol{m}}$ as 
\begin{equation}
\overset{0}{\boldsymbol{m}}=(m_1,m_2,\ldots,m_n,\underbrace{0,\ldots,0}_{n \:\:\text{entries}}),
\end{equation}
\begin{equation}
\overset{k}{\boldsymbol{m}}=(m_1,m_2,\ldots,m_n,0\ldots,\underbrace{m_{n+k}}_{\text{position n+k}},0,\ldots,0), \quad k=1,\ldots,n.
\end{equation}
Equipped with the above vector forms, we provide the following proposition.
\begin{proposition}
The moment of $\boldsymbol{x}(t)$ corresponding to $\boldsymbol{m}$ in state $k$ is non-zero if and only if $\boldsymbol{m}$ follows the corresponding form $\overset{k}{\boldsymbol{m}}$ for $k=0,\ldots,n$. Moreover, these moments verify the differential equations below
\begin{equation}
\frac{d\mu^{\overset{0}{\boldsymbol{m}}}_{0}(t)}{dt}=\sum_{i=1}^{n}m_i\mu^{\overset{0}{\boldsymbol{m}}-\boldsymbol{e}_i}_{0}(t)
-\sum_{i=1}^{n}a_i\mu^{\overset{0}{\boldsymbol{m}}+\boldsymbol{e}_i}_{0}(t)+\sum_{i=1}^{n}H_i\mu^{\overset{0}{\boldsymbol{m}}_{\overline{i}}}_{i}(t),
\label{zerostatemu}
\end{equation}
\begin{equation}
\frac{d\mu^{\overset{k}{\boldsymbol{m}}}_{k}(t)}{dt}= \begin{cases}
    a_k\mu^{\overset{k}{\boldsymbol{m}}+\boldsymbol{e}_k}_{0}(t)+\displaystyle\sum_{i=1}^{n}m_i\mu^{\overset{k}{\boldsymbol{m}}-\boldsymbol{e}_i}_{k}(t)-H_k\mu^{\overset{k}{\boldsymbol{m}}}_{k}(t)
    , \\ \textnormal{for } m_{n+k}=0,k\neq0,  \\
    \displaystyle\sum\limits_{\substack{i=1 }}^n m_i\mu^{\overset{k}{\boldsymbol{m}}-\boldsymbol{e}_i}_{k}(t)+m_{n+k}\mu^{\overset{k}{\boldsymbol{m}}-\boldsymbol{e}_{n+k}}_{k}(t)-H_k\mu^{\overset{k}{\boldsymbol{m}}}_{k}(t), \\ \textnormal{for } m_{n+k}\neq0,k\neq0, 
  \end{cases}
\end{equation}
where $\boldsymbol{e}_i\in\{0,1\}^{2n}$ is a unit vector with $1$ in position $i$ and $0$ elsewhere, and $\overset{0}{\boldsymbol{m}}_{\overline{i}}$ denotes the following vector
\begin{equation}
(m_1,m_2,\ldots,m_{i-1},0,m_{i+1},\ldots,m_n,0,\ldots,\underbrace{m_i}_{\text{at position i+n}},0,\ldots,0).
\end{equation}
\label{differentialcsma}
\end{proposition}
\begin{IEEEproof}
The first step of our analysis consists of understanding how the age evolves in each discrete state. To that end, we recall that in state $0$, there are no packets in the system. Accordingly, and as it has been previously explained, we have $x_{n+1}=x_{n+2}=\ldots=x_{2n}=0$ in this state. With that in mind, the vector $\boldsymbol{m}$ is required to have the following form 
\begin{equation}
\overset{0}{\boldsymbol{m}}=(m_1,m_2,\ldots,m_n,\underbrace{0,\ldots,0}_{n \:\:\text{entries}}),
\end{equation}
in order to ensure that the moment of $\boldsymbol{x}$ corresponding to the vector $\boldsymbol{m}$ in state $0$ is non-zero. By applying Theorem \ref{theoremillustrative} to this particular choice of $\boldsymbol{m}$, we get
\begin{equation}
\frac{d\mu^{\overset{0}{\boldsymbol{m}}}_{0}(t)}{dt}=\sum_{i=1}^{n}m_i\mu^{\overset{0}{\boldsymbol{m}}-\boldsymbol{e}_i}_{0}(t)
-\sum_{i=1}^{n}a_i\mu^{\overset{0}{\boldsymbol{m}}+\boldsymbol{e}_i}_{0}(t)+\sum_{i=1}^{n}H_i\mu^{\overset{0}{\boldsymbol{m}}_{\overline{i}}}_{i}(t),
\end{equation}
where $\boldsymbol{e}_i$ is a unit vector with $1$ in position $i$ and $0$ elsewhere, and $\overset{0}{\boldsymbol{m}}_{\overline{i}}$ denotes the following vector
\begin{equation}
(m_1,m_2,\ldots,m_{i-1},0,m_{i+1},\ldots,m_n,0,\ldots,\underbrace{m_i}_{\text{position i+n}},0,\ldots,0). 
\end{equation}
In other words, in the vector above, the position $i$ becomes equal to zero, and the position $i+n$ inherits the value $m_i$. This is a consequence of the fact that a transition from state $i$ to state $0$ happens when a packet is delivered. When such an event happens, the age at the monitor $x_i$ becomes equal to the age of the delivered packet $x_{i+n}$. \color{black}To make our results clearer, we detail in the following how we applied Theorem \ref{theoremillustrative}. To that end, we adopt the test function 
$\psi_0^{\overset{0}{\boldsymbol{m}}}(q(t),\boldsymbol{x},t)=\boldsymbol{x}^{\overset{0}{\boldsymbol{m}}}(t)\delta_{0q(t)}$. By applying Theorem \ref{theoremillustrative} while keeping in mind the transitions dynamics, we obtain

\begin{align}
&L\psi^{\overset{0}{\boldsymbol{m}}}_{0}(q(t),\boldsymbol{x},t)=\delta_{0q(t)}[\sum_{i=1}^{n}[m_ix_i^{m_i-1}\prod_{j=1,j\neq i}^{n}
 x_j^{m_j}]\nonumber\\&-\sum_{i=1}^{n}a_ix_i\boldsymbol{x}^{\boldsymbol{m}}]+\sum_{i=1}^{n}\delta_{iq(t)}H_i\boldsymbol{x}^{\overset{0}{\boldsymbol{m}}_{\overline{i}}}_{i}.
\end{align}
Then, given the definition of $\mu^{\boldsymbol{m}}_{\overline{q}}(t)$ provided in eq. (\ref{mumstuff}) for $\overline{q}\in\mathbb{Q}$ and the equality in (\ref{derivativestufff}), we can obtain the results of (\ref{zerostatemu}). 
\color{black}

%
%
%
Next, we study the moment of $\boldsymbol{x}$ with respect to $\boldsymbol{m}$ in any other discrete state $k\neq0$. We point out that if the discrete process $q(t)$ is equal to $k\neq0$, then there exists a packet for link $k$ in the system, but there are no packets for the remaining links. Accordingly, $x_{i}=0$ for $i\in\{n+1,\ldots,2n\}\setminus\{k\}$. Therefore, the vector $\boldsymbol{m}$ is required to have the following form 
\begin{equation}
\overset{k}{\boldsymbol{m}}=(m_1,m_2,\ldots,m_n,0\ldots,\underbrace{m_{n+k}}_{\text{position n+k}},0,\ldots,0),
\end{equation}
in order to ensure that the moment of $\boldsymbol{x}$ corresponding to the vector $\boldsymbol{m}$ in state $k$ is non-zero. By applying Theorem \ref{theoremillustrative} for this particular choice of $\boldsymbol{m}$, we get
\begin{equation}
\frac{d\mu^{\overset{k}{\boldsymbol{m}}}_{k}(t)}{dt}= \begin{cases}
    a_k\mu^{\overset{k}{\boldsymbol{m}}+\boldsymbol{e}_k}_{0}(t)+\displaystyle\sum_{i=1}^{n}m_i\mu^{\overset{k}{\boldsymbol{m}}-\boldsymbol{e}_i}_{k}(t)-H_k\mu^{\overset{k}{\boldsymbol{m}}}_{k}(t)
    , \nonumber\\ \textnormal{for } m_{n+k}=0,k\neq0,  \\
    \displaystyle\sum\limits_{\substack{i=1}}^n m_i\mu^{\overset{k}{\boldsymbol{m}}-\boldsymbol{e}_i}_{k}(t)+m_{n+k}\mu^{\overset{k}{\boldsymbol{m}}-\boldsymbol{e}_{n+k}}_{k}(t)-H_k\mu^{\overset{k}{\boldsymbol{m}}}_{k}(t), \nonumber\\ \textnormal{for } m_{n+k}\neq0,k\neq0, 
  \end{cases}
\end{equation}
which concludes our proof. 
\end{IEEEproof}

Given the positive recurrence of the process, along with the Lagrange stability that eliminates the possibility of infinite moments, we can be assured that as time passes by, we have
\begin{equation}
\boldsymbol{\mu}^{\boldsymbol{m}}(t)\xrightarrow{t\rightarrow\infty}\boldsymbol{\mu}^{\boldsymbol{m}}_{\infty},
\end{equation}
where $\boldsymbol{\mu}^{\boldsymbol{m}}_{\infty}$ is the steady-state moment of order $\boldsymbol{m}$. 
Given the ODEs depicted in Proposition \ref{differentialcsma}, we can stack the equations that the moments verify for any order $\boldsymbol{m}\geq0$ and take their derivative to zero. Accordingly, we get that the vector $\boldsymbol{\mu}_{\infty}$ verifies the following linear system
\begin{equation}
\boldsymbol{A}_{\infty}\boldsymbol{\mu}_{\infty}=\boldsymbol{b}_{\infty},
\end{equation}
where 
\begin{equation}
\boldsymbol{\mu}_{\infty}=[\mu^{\boldsymbol{0}}_0(\infty), \mu^{\boldsymbol{0}}_1(\infty),\ldots,\mu^{\boldsymbol{0}}_{n}(\infty),\mu^{\boldsymbol{e}_1}_{0}(\infty),\ldots],
\end{equation}
and $\boldsymbol{A}_{\infty}$ and $\boldsymbol{b}_{\infty}$ are an infinite dimension matrix and vector respectively that incorporate the entries of the ODE for every $\boldsymbol{m}\geq0$. Therefore, the remaining step of our analysis is to solve the above system and find an expression of the first-order moment of the age process of interest.
\section{Solutions to the SHS}
\label{shssolutions}
\subsection{Moment Closure Method}
In both systems depicted in Section \ref{bassingle} and \ref{bassmultiple}, we have shown that the steady-state moments can be computed by solving the following infinite-dimensional system of linear equations
\begin{equation}
\boldsymbol{A}_{\infty}\boldsymbol{\mu}_{\infty}=\boldsymbol{b}_{\infty}.
\end{equation}
However, the fact remains that any moment of a particular order will depend on higher-order moments and so on (hence the infinite dimension aspect of the above system). This infinite aspect of the system renders solving the above linear system impossible. Researchers have studied this type of system extensively as they are prevalent in many applications such as chemical kinetics \cite{pmid21338614}, physics \cite{FryNov1982}, population dynamics/epidemiology \cite{doi:https://doi.org/10.1002/9781444311501.ch3}. Concretely, the primary approach in the literature to address this issue is referred to as the \emph{moment closure} technique. To understand this method, let us suppose that we are only interested in the dynamics of moments up to a particular order $k$. To that end, we can summarize the steady-state dynamics of these moments as follows
\begin{equation}
\boldsymbol{A}_{\infty|k}\boldsymbol{\mu}_{\infty|k}+\boldsymbol{B}\overline{\boldsymbol{\mu}}_{\infty}=\boldsymbol{b}_{\infty|k},
\label{redefinedsmallsystem}
\end{equation}
where 1) $\boldsymbol{\mu}_{\infty|k}$ denotes the moments up till order $k$, 2) $\overline{\boldsymbol{\mu}}_{\infty}$ refers to the set of moments of order higher than $k$ that $\boldsymbol{\mu}_{\infty|k}$ depends on, 3) $\boldsymbol{A}_{\infty|k}$ denotes the section of $\boldsymbol{A}_{\infty}$ that relate the moments $\boldsymbol{\mu}_{\infty|k}$ to one another, 4) $\boldsymbol{B}$ depicts the relationship between the moments $\boldsymbol{\mu}_{\infty|k}$ and $\overline{\boldsymbol{\mu}}_{\infty}$, and 5) $\boldsymbol{b}_{\infty|k}$ consists of the section of $\boldsymbol{b}_{\infty|k}$ relevant to $\boldsymbol{\mu}_{\infty|k}$. As can be seen, the main issue from here on is to find the best way to approximate the higher-order moments $\overline{\boldsymbol{\mu}}_{\infty}$. By doing so, the system becomes closed and fully determined, and one can then find the solution $\boldsymbol{\mu}_{\infty|k}$ of the finite linear system. To carry out this approximation, several methods have been proposed in the literature
\begin{itemize}
\item One of the most straightforward approaches is to ignore the higher-order moments and suppose that all the components of $\overline{\boldsymbol{\mu}}_{\infty}$ are equal to zero \cite{Socha_2008}. 
\item Another technique consists of writing the components of $\overline{\boldsymbol{\mu}}_{\infty}$ as non-linear functions of $\boldsymbol{\mu}_{\infty|k}$. An example of this technique consists of adopting a derivatives matching method to find this non-linear function form \cite{Hespanha2007ModelingAA}. 
\item In a different line of work, researchers assumed a specific distribution for the higher-order moments, which can allow us to close the system. The examples of such approaches range from simple Gaussian distribution assumptions as done by Whittle back in 1957 \cite{10.2307/2983819} to more elaborate methods such as entropy maximization recently proposed in \cite{7039471}. 
\end{itemize}
We refer the readers to \cite{Kuehn2016} for a thorough review on the moment closure techniques proposed in the literature. In the sequel, we will provide a step-by-step motivation for the approach that we will adopt in our framework. 

The best way to understand our approach is for us to go back to the simple example we have considered back in Section \ref{depiction}. Given the ODE previously reported in eq. (\ref{odeillustrative}), and by taking the derivatives to zero in the steady-state regime, the infinite-dimensional system of interest is $\boldsymbol{A}_{\infty}\boldsymbol{\mu}_{\infty}=\boldsymbol{b}_{\infty}$ where
\begin{equation*}
\boldsymbol{\mu}_{\infty}=\begin{bmatrix}
           \mu^1_{\infty} \\
           \mu^2_{\infty} \\
          \vdots \\
         \end{bmatrix} 
         \boldsymbol{b}_{\infty}=\begin{bmatrix}
           -1 \\
           0 \\
          \vdots \\
         \end{bmatrix} 
\end{equation*}
\begin{equation}
\boldsymbol{A}_{\infty}=\begin{bmatrix}
0 & -a_1 & 0 & \cdots  & \cdots &   \cdots\\
2 & 0 & -a_1 & 0  & \ddots  & \cdots\\
0 & 3 & \ddots & \ddots & \ddots & \ddots\\
0 & 0 & \ddots & \ddots & \ddots & \ddots\\
\end{bmatrix}
\end{equation}

Let us suppose that we are only interested in considering the steady-state dynamics of the age moments up till order $4$. By proceeding with a moment closure approach, we can summarize the dynamics of these moments by using eq. (\ref{redefinedsmallsystem}) with 
\begin{align}
&\boldsymbol{\mu}_{\infty|4}=\begin{bmatrix}
           \mu^1_{\infty} \\
           \mu^2_{\infty} \\
          \mu^3_{\infty} \\
           \mu^4_{\infty} 
         \end{bmatrix} 
         \boldsymbol{b}_{\infty|4}=
\begin{bmatrix}
           -1 \\
           0 \\
          0 \\
           0 
         \end{bmatrix}
\boldsymbol{A}_{\infty|4}=         
         \begin{bmatrix}
           0 & -a_1 & 0 & 0 \\
           2 & 0 & -a_1 & 0\\
          0 & 3 & 0 & -a_1\\
          0 & 0 & 4 & 0     
         \end{bmatrix}         
         \nonumber\\&
         \boldsymbol{B}=\begin{bmatrix}
           0 & 0 & 0 & 0 & 0 \\
            0 & 0 & 0 & 0 & 0\\
           0 & 0 & 0 & 0 & 0\\
            0 & 0 & 0 & 0 & -a_1     
         \end{bmatrix} 
         \overline{\boldsymbol{\mu}}_{\infty}=\begin{bmatrix}
           0 \\
           0 \\
          0\\
          0 \\
           \mu^5_{\infty}
         \end{bmatrix}
\end{align}
To motivate our proposed approach, let us consider that the value of $a_1$ is high. In this regime, the transition of the age process back to zero happens at a significantly high rate. Therefore, we can expect the age process $x(t)$ to be generally small. Specifically, we expect the age process to be smaller than $1$ with a high probability. This itself will lead to a small first-order moment. Given that the function $f(m)=c^m$ is decreasing when $c<1$, we can also expect higher-order moments to be even smaller. For example, by fixing $a_1$ to $100$, we obtain through Monte Carlo simulations the following steady-state moments
\begin{center}
\begin{tabular}{|c|c|c|}
 \hline
 Order & Moment Value \\
  \hline
 $1$  & $0.0785$ \\
 $2$   & $0.01$ \\
 $3$   & $0.0016$ \\
 \hline
\end{tabular}
 \captionof{table}{Age moments in the illustrative example for $a_1=100$. }
 \label{RRCSMA}
\end{center}
With this in mind, let us focus our attention on the moments up till order $m$. To close the linear system, we suppose that the moment of order $m+1$ is equal to the moment of order $m$. This is motivated by the fact that as $m$ grows, both these moments will be significantly small to the point that the difference between them can be neglected. Therefore, the closure of the system will come at a minor penalty. To illustrate this, we plot below the curve of the estimate of the average age (i.e., $\mu^1_{\infty}$) using our closure method when we vary $m$.
\begin{figure}[!ht]
\centering
\includegraphics[width=.65\linewidth]{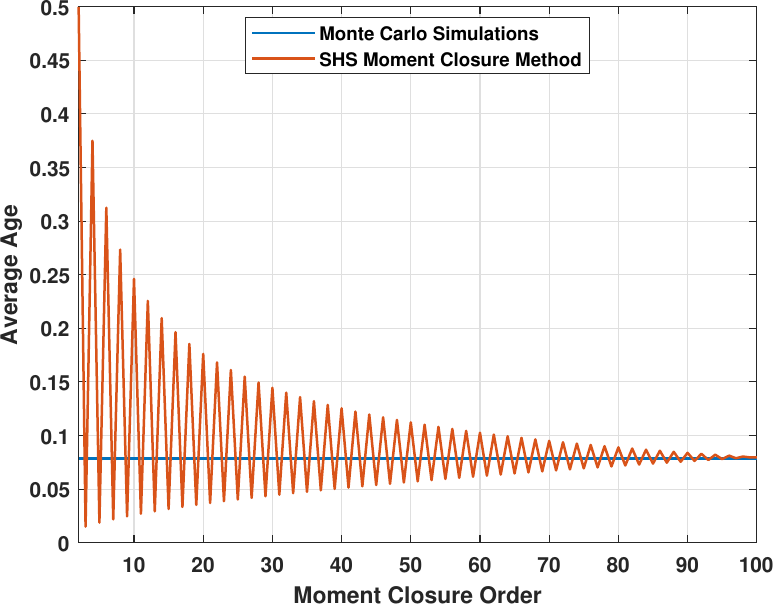}
\setlength{\belowcaptionskip}{-5pt}
\caption{Implementation of the proposed moment closure method for $a_1=100$.}
\label{illustrationexampless}
\end{figure}\\
As can be seen in Fig. \ref{illustrationexampless}, by adopting our proposed method, we can approach the actual value of the average age by closing the system at a high $m$\footnote{Note that the fluctuations originate from the differences between closing the system at an odd or even value of $m$.}. Getting back to our linear system, we can conclude that it is sufficient to solve a linear system of dimension $100$ to get an accurate estimate of the average age. 

Now, let us suppose that we tune $a_1$ back to a small value. In this case, the transitions that set the age back to zero happen infrequently. We can therefore expect the age to be larger than $1$ with a high probability. Accordingly, we cannot proceed in the same way since the higher-order moments will be significantly large, and therefore, approximating the moment of order $m+1$ by that of order $m$ will be highly inaccurate. For example, by fixing $a_1$ to $0.1$, we can obtain through Monte Carlo simulations the following steady-state moments
\begin{center}
\begin{tabular}{|c|c|c|}
 \hline
 Order & Moment Value \\
  \hline
 $1$  & $2.55$ \\
 $2$   & $10.27$ \\
 $3$   & $52.4$ \\
 \hline
\end{tabular}
 \captionof{table}{Age moments in the illustrative example for $a_1=0.1$. }
 \label{RRCSMA12}
\end{center}
Let us now see how our proposed approach works in this case. To that end, we plot below the curve of the estimate of the average age using our closure method when we vary $m$.
\begin{figure}[!ht]
\centering
\includegraphics[width=.65\linewidth]{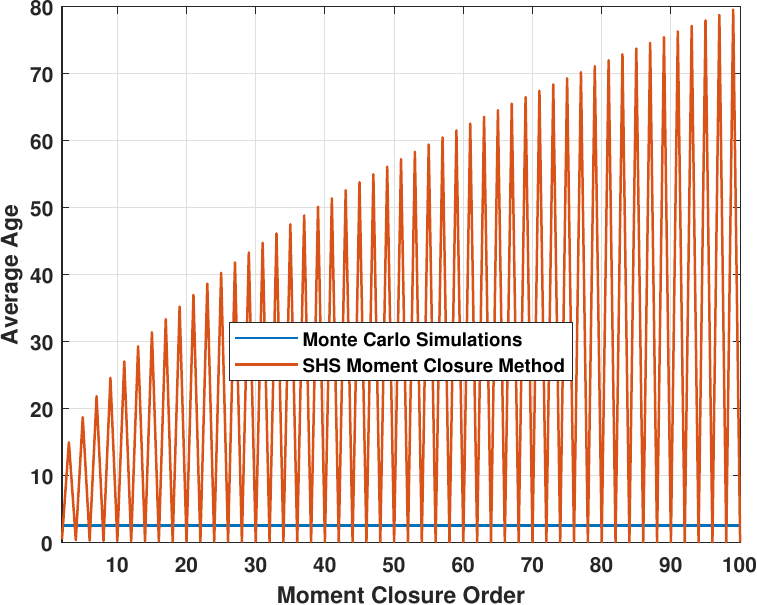}
\setlength{\belowcaptionskip}{-5pt}
\caption{Implementation of the proposed moment closure method for $a_1=0.1$.}
\label{illustrationexamplessbaddaaa}
\end{figure}\\
It can be easily seen from the results in Fig. \ref{illustrationexamplessbaddaaa} that our approach fails in this case. Therefore, how can we use our proposed method when the age processes are not generally small? To solve this issue, we proceed by tweaking the differential equations reported in eq. (\ref{odeillustrative}). Specifically, instead of following the evolution of the age process $x(t)$, we track the evolution of a scaled process $z(t)=x(t)/c$ where $c\geq1$ is a large fixed number. In this case, the ODE will become
\begin{equation}
\frac{c^m d\mathbb{E}[z^m(t)]}{dt}=mc^{m-1}\mathbb{E}[z^{m-1}(t)]-a_1c^{m+1}\mathbb{E}[z^{m+1}(t)].
\label{newodeillustrative}
\end{equation}
By multiplying the whole equation by $1/c^{m-1}$, and by taking the left-hand side equal to zero, we can proceed similarly as before to obtain a linear system for us to solve for the scaled process. We can then close the infinite linear system up to an order $m$. Given that $c$ is large, we can then use the same method of approximating the moment of order $m+1$ by the moment of order $m$. Afterward, we can deduce the steady-state first-order moment $\mathbb{E}[z^1(\infty)]$ and scale it back up by multiplying it by $c$ to conclude the average age of the original process. The results of this approach are reported in the figure below.
\begin{figure}[!ht]
\centering
\includegraphics[width=.65\linewidth]{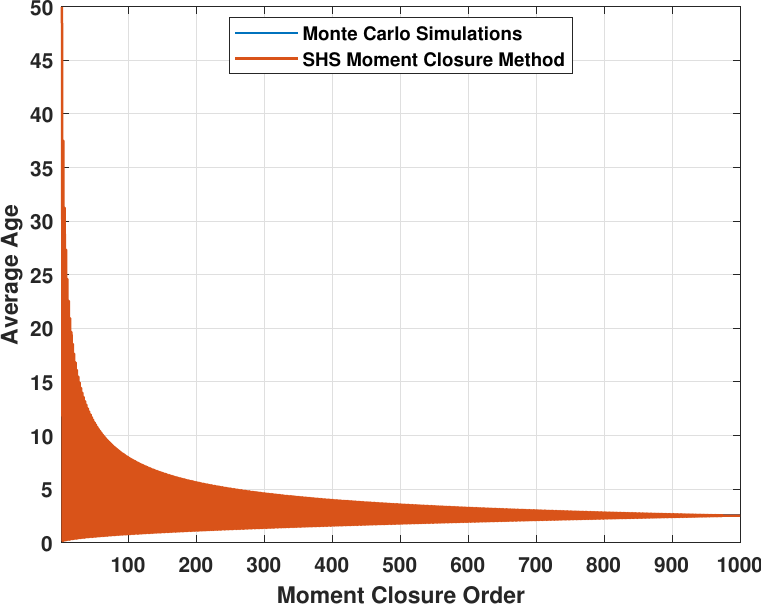}
\setlength{\belowcaptionskip}{-5pt}
\caption{Implementation of the scaled random variable approach for $a_1=0.1$.}
\label{illustrationexamplessfixed}
\end{figure}\\
As we can see in Fig. \ref{illustrationexamplessfixed}, we were able to avoid the issues that arose in Fig. \ref{illustrationexamplessbaddaaa} and approximate with precision the average age of interest. Note that this same approach can be adopted for the multiple processes case by scaling each individual age process using the large fixed number $c\geq1$. In essence, our approach consists of scaling the ODE by a sufficiently large constant $c\geq1$ when necessary and simply approximate the high order moments by lower-order ones in this regime. As discussed, the literature on moment closures techniques is rich, and many new methods are being proposed to this day. However, given that this method performs well, as seen in the above example, and given its simplicity and ease of implementation, we will adopt it in the rest of this paper. 

\begin{remark}
The choice of the moment order $m^*$ beyond which all higher moments may be ignored will depend on the system in question. Note that based on Theorem \ref{theoremillustrative}, we can see that the dependency between the moments of any order $m$ on higher-order moments is contingent on the polynomial degree of $\lambda_l(\cdot)$. Accordingly, we can see that the resulting matrix of the linear system that we wish to solve is typically very sparse. Therefore, the complexity of solving the linear system can be reduced even for large values of $m$ \cite{doi:10.1137/1.9781611976465.31}. With this in mind, one can proceed as follows: 1) define a grid of moments order cut off $[m_0,m_1,\ldots,m_{i_{\textnormal{max}}}]$ with a predefined step size $M=m_{i+1}-m_{i}$, 2) solve the linear system in ascending order until the incorporation of a larger set of high order moments does not lead to a significant increase in the accuracy of the average AoI. In other words, one can take $m^*=m_i$ once the average AoI estimation at $m_i$ and $m_{i+1}$ are within a certain predefined threshold $\epsilon$.
\end{remark}
\color{black}
\subsection{Parameters Optimization}
\label{parametrssoptimization}
The above moment closure technique has provided us with a simple way to compute the average age of any system modeled through the SHSs tool depicted earlier. Specifically, we were able to show that by scaling the ODE using a sufficiently large constant $c\geq1$, and by solving a linear system of a sufficiently high dimension $m$, we can obtain an approximation of the average age as follows
\begin{equation}
\overline{\Delta}\backsimeq c\boldsymbol{d}^T\boldsymbol{\mu}_{\infty|k}=c\boldsymbol{d}^T\boldsymbol{E}^{-1}\boldsymbol{b}_{\infty|m},
\label{Nusersopt}
\end{equation}
where the matrix $\boldsymbol{E}$ takes into account $\boldsymbol{A}_{\infty}$, $\boldsymbol{B}$, and our closure technique of the higher-order moments of the scaled ODE. On the other hand, the vector $\boldsymbol{d}$ allows us to extract the first-order components of the moments vector that we are interested in. Now, let us go further than this and try to optimize the average age by calibrating the transitions rates parameters to achieve the best possible age performance. Specifically, let us suppose that there exists a set of parameters $\boldsymbol{\eta}\in\mathcal{X}$ that can be controlled by the designer, where $\mathcal{X}$ is a compact set. For example, in the illustrative example reported in Section \ref{depiction}, we have $\eta=a_1$. Our goal becomes to find the optimal $\boldsymbol{\eta}^*$ to minimize the average age. In other words, we can formulate our optimization problem as follows
\begin{equation}
\begin{aligned}
& \underset{\boldsymbol{\boldsymbol{\eta}}\in\mathcal{X}}{\text{minimize}}
&& \overline{\Delta}(\boldsymbol{\eta})=c\boldsymbol{d}^T\boldsymbol{E}^{-1}(\boldsymbol{\eta})\boldsymbol{b}_{\infty|m}(\boldsymbol{\eta})\\
\end{aligned}
\label{Nusersopt11}
\end{equation}
To solve the above problem, we note that finding a closed-form of the objective function in (\ref{Nusersopt11}) is not always feasible. Therefore, we circumvent this difficulty by employing a sequential convex approximation approach. To that end, the proposed SCA approach can be summarized below
\begin{equation}
\hat{\boldsymbol{\eta}}[k]= \underset{\boldsymbol{\eta}\in \mathcal{X}}{\text{argmin}} \:\:\overline{\Upsilon}(\boldsymbol{\eta},\overline{\boldsymbol{\eta}}[k]), \quad k=1,2\ldots,
\label{scaprocedure}
\end{equation}
\vspace{-2pt}
where
\begin{equation}
\overline{\Upsilon}(\boldsymbol{\eta},\overline{\boldsymbol{\eta}}[k])=\overline{\Delta}(\overline{\boldsymbol{\eta}}[k])+\nabla\overline{\Delta}(\overline{\boldsymbol{\eta}}[k])^{T}(\boldsymbol{\eta}-\overline{\boldsymbol{\eta}}[k])+\frac{1}{2\alpha_k}||\boldsymbol{\eta}-\overline{\boldsymbol{\eta}}[k]||^2_{2}.
\label{scafunction}
\end{equation}
The term $\frac{1}{2\alpha_k}||\boldsymbol{\eta}-\overline{\boldsymbol{\eta}}[k]||^2_{2}$ is a regularization term that is employed to keep the points close enough so that the model is accurate. We will report conditions on $\alpha_k$ to ensure the convergence of the approach in Proposition \ref{finalconvergence}. Since the problem in (\ref{scaprocedure}) is convex, we can solve it using standard convex solvers such as CVX \cite{cvx}. After finding the solution of (\ref{scaprocedure}), at each iteration, we set $\overline{\boldsymbol{\eta}}[k+1]=\hat{\boldsymbol{\eta}}[k]$. The next step consists of finding the expression of the gradient of the average age. To do so, we observe that 
\begin{equation}
\frac{\partial \overline{\Delta}(\boldsymbol{\eta})}{\partial \eta_{i}}=c\boldsymbol{d}^T\frac{\partial \boldsymbol{E}^{-1}(\boldsymbol{\eta})}{\partial \eta_{i}}\boldsymbol{b}_{\infty|m}(\boldsymbol{\eta})+c\boldsymbol{d}^T\boldsymbol{E}^{-1}(\boldsymbol{\eta})\frac{\partial \boldsymbol{b}_{\infty|m}(\boldsymbol{\eta})}{\partial \eta_{i}}.
\end{equation}
Using the following identity
\begin{equation}
\frac{\partial \boldsymbol{K}^{-1}}{\partial x_{i}}=-\boldsymbol{K}^{-1}\frac{\partial \boldsymbol{K}}{\partial x_{i}}\boldsymbol{K}^{-1},
\end{equation}
we can conclude that
\begin{align}
\frac{\partial \overline{\Delta}(\boldsymbol{\eta})}{\partial \eta_{i}}=&-c\boldsymbol{d}^T\boldsymbol{E}^{-1}(\boldsymbol{\eta})\frac{\partial \boldsymbol{E}(\boldsymbol{\eta})}{\partial \eta_{i}}\boldsymbol{E}^{-1}(\boldsymbol{\eta})\boldsymbol{b}_{\infty|m}(\boldsymbol{\eta})\nonumber\\&+c\boldsymbol{d}^T\boldsymbol{E}^{-1}(\boldsymbol{\eta})\frac{\partial \boldsymbol{b}_{\infty|m}(\boldsymbol{\eta})}{\partial \eta_{i}}
\end{align}
Based on the above equation, the gradient vector $\nabla\overline{\Delta}(\boldsymbol{\eta})$ can be found. We summarize our approach in Algorithm 1.
\begin{algorithm}
\caption{Proposed SCA approach}\label{euclid}
\begin{algorithmic}[1]
\State \textbf{Input} Stopping criterion $\epsilon$ and two feasible points $\overline{\boldsymbol{\eta}}[1],\hat{\boldsymbol{\eta}}[1] \in \mathcal{X}$
\State \textbf{Initialize} Set $k=1$
\State \textbf{Iterate}
\State $\overline{\boldsymbol{\eta}}[k+1]:=\hat{\boldsymbol{\eta}}[k]$
\State $k:=k+1$
\State Solve the convex problem in (\ref{scaprocedure}) to find $\hat{\boldsymbol{\eta}}[k]$
\State \textbf{Until} $||\overline{\Upsilon}(\hat{\boldsymbol{\eta}}[k],\overline{\boldsymbol{\eta}}[k])-\overline{\Upsilon}(\hat{\boldsymbol{\eta}}[k-1],\overline{\boldsymbol{\eta}}[k-1])||<\epsilon$
\State Output $\overline{\boldsymbol{\eta}}[k]$
\end{algorithmic}
\end{algorithm}\\
In the sequel, we provide a convergence analysis of the Algorithm presented above. To proceed in that direction, we first lay out the following definition.
\begin{definition}[Stationary points of a function]
Let $f \colon \mathcal{D} \to \mathbb{R}$ be a function where $\mathcal{D}\subseteq\mathbb{R}^n$ is a convex set. A point $\boldsymbol{x}^*\in\mathcal{D}$ is a stationary point of $f(.)$ if $\nabla_{\vec{\boldsymbol{d}}}f(\boldsymbol{x}^*)\geq0$ for all $\vec{\boldsymbol{d}}\in\mathcal{D}$ such that $\boldsymbol{x}^*+\vec{\boldsymbol{d}}\in\mathcal{D}$. 
\end{definition}
\noindent Equipped with the above definition, we present the following convergence results.
\begin{proposition}
The sequence $\big\{\overline{\Upsilon}(\hat{\boldsymbol{\eta}}[k],\overline{\boldsymbol{\eta}}[k])\big\}_{k=1}^{+\infty}$ is convergent for $\alpha_k\geq \frac{L}{2},\:k\in\mathbb{N}$ where $L$ is the Lipschitz constant of the function $\nabla\overline{\Delta}(\boldsymbol{\eta})$. Moreover, the limit point of the sequence $\big\{\overline{\boldsymbol{\eta}}[k]\big\}_{k=1}^{+\infty}$ generated by the SCA procedure (\ref{scaprocedure}) is a stationary point of the problem in (\ref{Nusersopt11}).
\label{finalconvergence}
\end{proposition}
\begin{IEEEproof}
The proof can be found in Appendix \ref{appendixfinalconvergence}.
\end{IEEEproof}
\noindent Therefore, the SCA algorithm provided above guarantees that we obtain a stationary point $\boldsymbol{\eta}^*$ of the average age. 
\section{Numerical Implementations}
\label{numericalimplementations}
With our complete theoretical analysis laid out, this section aims to showcase the usefulness of the SHSs tool provided in this paper. To that end, we implement the age-aware CSMA scheme depicted in Section \ref{environmentdescrip}. By leveraging the results of the SHS analysis presented in our paper, we optimize this age-aware CSMA scheme using the SCA approach detailed in Section \ref{parametrssoptimization}. As a benchmark, we consider a CSMA scheme where users employ a constant back-off rate when accessing the channel. In this case, the AoI-optimal back-off rate was previously found in \cite{9007478}. Note that it is evident that the age-aware CSMA scheme offers more degrees of freedom and, therefore, has the potential of improving the AoI compared to the latter approach. Given that our SHS analysis will allow us to extract this performance improvement, the question that remains is how significant this gain in performance is. To put this into perspective, let us consider a $2$ users scenario with $H_1$ fixed to $1$ and $H_2$ being variable. To obtain the best comparison between the two schemes, we implement them in a realistic IEEE 802.11 environment and plot the performance gain of the age-dependent CSMA compared to the traditional age-optimal CSMA in function of $H_2$.\color{black}
\begin{figure}[!ht]
\centering
\includegraphics[width=.75\linewidth]{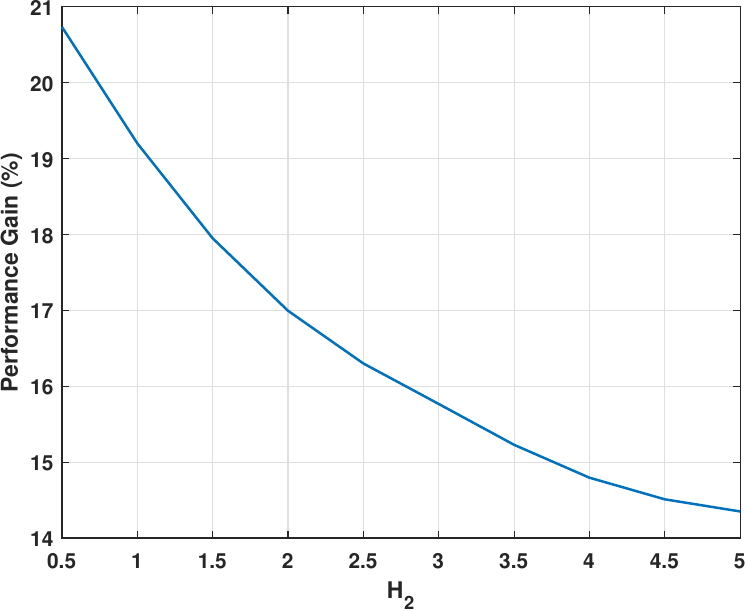}
\setlength{\belowcaptionskip}{-5pt}
\caption{Performance gain between age-aware and age-blind CSMA.}
\label{CSMAimplementation}
\end{figure}\\
As shown in Fig. \ref{CSMAimplementation}, the age-aware CSMA can reduce the average age by as much as $20\%$ compared to the optimal age-blind approach. This puts into perspective the usefulness of the SHSs tool analysis provided in this paper as it opens up the gate to various new applications/systems to be studied and performance gains to be revealed. 
\section{Conclusion}
\label{conclusionss}
In this paper, we have studied a general status update system modeled using the SHSs tool where the system's transition dynamics are allowed to be functions of the AoI. To analyze this system, we have shown several critical aspects that the AoI-dependent SHS model verifies. We have also proved the Lagrange stability of the ODEs involved, along with the positive recurrence of the age processes. Additionally, we have proposed a moment closure technique that allows us to compute the average age of the system. Equipped with this method, we have also provided an SCA approach to optimize the system's parameters. Finally, an age-dependent CSMA system was implemented to showcase the performance advantage that the age-dependency provides. To that end, the results of this paper provide a framework that will allow the study of more elaborate and complex status update systems in the future.
\bibliographystyle{IEEEtran}
\bibliography{trialout}
\appendices
\section{Proof of Theorem \ref{stabilitysinglee}}
\label{appendixstabilitysingle}
To proceed with our proof, let us start with some useful definitions. To that end, let $J_{\text{max}}=\underset{l}{\text{max}}\:J_l$ be the maximum degree of the transitions rates, and $l^*$ be the index such that $J_{l^*}=J_{\text{max}}$. Now let us suppose that $\mu^{m}(t)$ is arbitrarily large. In other words, for any $\epsilon>0$, we have
\begin{equation}
\mu^{m}(t)\geq \frac{1}{\epsilon}.
\end{equation}
Given the Jensen's inequality results reported in eq. (\ref{illustrumomentcomparisonconvex}), we can conclude that
\begin{equation}
\mu^{m+J_{\text{max}}}(t)\geq [\mu^{m}(t)]^{(m+J_{\text{max}})/m}\geq \frac{\mu^{m}(t)}{\epsilon^{J_{\text{max}}/m}}.
\label{relatingthem}
\end{equation}
The above inequality tells us that if the moment of order $m$ is arbitrarily large, then the moment of order $m+J_{\text{max}}$ grows at an even larger pace. If $J_{\text{max}}=0$, then we go back to the case studied in \cite{8469047} where the stability was established, and the finiteness can be consequently obtained. More generally, when $J_{\text{max}}\geq1$, we note that the moment of order $m+J_{\text{max}}$ has a strictly negative coefficient in eq. (\ref{singlestatespecific}). Given that a negative coefficient is tied to the highest order moment (i.e., $\mu^{m+J_{\text{max}}}(t)$), and the inequality provided in eq. (\ref{relatingthem}), we can assert that there exists a certain $\epsilon_0$ such that for all $\epsilon\leq\epsilon_0$, we have
\begin{equation}
\frac{d\mu^m(t)}{dt}<0.
\end{equation}
In other words, as $\mu^m(t)$ grows larger, eventually the derivative of $\mu^m(t)$ becomes negative and the value of the moment of order $m$ is pulled back towards the origin. Accordingly, we can conclude that there exists a large enough ball centered around the origin of radius $U_m$ such that $\mu^m(t)\in B(0,U_m)$ for $t\in[0,\infty)$.
\section{Proof of Proposition \ref{exitfinitesingle}}
\label{proofexitfinemnawalwahad}
Let us consider the following twice differentiable function
\begin{equation}
W^{m}(x)=k-(x+1)^m, \quad \forall x\in D,
\end{equation}
where the constants $k$ and $m$ are to be specified. We apply the SHS extended generator previously reported in eq. (\ref{shsextendedgenerator}) on this function $W^{m}(x)$. We end up with
\begin{equation}
LW^{m}(x)=-m(x+1)^{m-1}+\sum_{l=1}^{L}[(x+1)^m-(c+1)^m][\sum_{j=0}^{J_{l}}a^{(l)}_jx^j].
\end{equation}
Given that $D$ has a compact closure, we can deduce that there exists $U\geq1$ such that
\begin{equation}
1\leq x+1\leq U.
\label{inequalitiesonxxx}
\end{equation}
Based on the above, we can also conclude that there exists $U^*\geq0$ such that
\begin{equation}
\sum_{j=0}^{J_{l}}a^{(l)}_jx^j\leq U^*, \quad l=1,\ldots,L.
\end{equation}
By keeping the above inequality in mind, and given that $c\geq0$, we can deduce that
\begin{align}
LW^{m}(x)&\leq-m(x+1)^{m-1}+((x+1)^m-1)LU^*\nonumber\\&=(x+1)^{m-1}(-m+(x+1)LU^*)-LU^*\nonumber\\
&\leq (x+1)^{m-1}(-m+ULU^*)-LU^*
\end{align}
Therefore, it sufficient to choose $m>ULU^*$ for the above Right Hand Side (\textbf{RHS}) to be strictly negative. In this case, there exists a sufficiently small $\epsilon>0$ such that
\begin{equation}
LW^{m}(x)\leq-\epsilon.
\end{equation}
Lastly, given eq. (\ref{inequalitiesonxxx}), we can easily choose a constant $k>0$ such that $W^{m}(x)\geq0$ for any $x\in D$. All in all, we can conclude that $W^{m}(x)$ is a Lyapunov function. Let us now define $\overline{\tau}_D(t)=\min\{t,\tau_D\}$. Given that Dynkin's formula is applicable (see \cite[Theorem~1]{HESPANHA20051353}), we obtain
\begin{align}
\mathbb{E}_{x}[W^{m}(x(\overline{\tau}_D(t))]-W^{m}(x)&=\mathbb{E}_{x}[\int_{0}^{\overline{\tau}_D(t)}LW^{m}(x(u))du]\nonumber\\&\leq -\epsilon\mathbb{E}_{x}[\overline{\tau}_D(t)].
\end{align}
Given that $W^{m}(\cdot)$ is non-negative, we get
\begin{equation}
\mathbb{E}_x[\overline{\tau}_D(t)]\leq\frac{W^{m}(x)}{\epsilon}.
\label{proofeqlowerbound}
\end{equation}
Next, we note that
\begin{equation}
\mathbb{E}_x[\overline{\tau}_D(t)]=\mathbb{E}_x[\tau_D\mathbbm{1}\{\tau_D<t\}]+\mathbb{E}_x[t\mathbbm{1}\{\tau_D>t\}].
\end{equation}
Given the above equation, along with eq. (\ref{proofeqlowerbound}), we can conclude that
\begin{equation}
t{\Pr}_{x}(\tau_D>t)\leq\frac{W^{m}(x)}{\epsilon}.
\end{equation}
Letting $t\rightarrow\infty$, we obtain
\begin{equation}
{\Pr}_{x}(\tau_D=\infty)=0.
\end{equation}
This tells us that $\overline{\tau}_D(t)\rightarrow\tau_D$ almost surely when $t\rightarrow\infty$. Now, by applying Fatou's lemma, we obtain 
\begin{equation}
\mathbb{E}_{x}[\tau_D]\leq \frac{W^{m}(x)}{\epsilon}<\infty.
\end{equation}
\color{black}
\section{Proof of Theorem \ref{positiverecurrencesinglestuff}}
\label{proofpositiverecurrencesinglestuff}
To prove our desired results, we first note that any non-empty open set $D\subset[c,\infty]$ with compact closure $\overline{D}$ can be written as $]\alpha,\beta[$ for some $\alpha>c$ and $\beta<\infty$. \color{black}Next, by leveraging the results of Proposition \ref{exitfinitesingle}, we know that we exit the set $D$ and enter $D^c$ in finite time. To that end, let us consider the two sets $D_{\textnormal{inf}}=[c,\alpha]$ and $D_{\textnormal{sup}}=[\beta,\infty)$. It is evident that $D^c=D_{\textnormal{inf}}\cup D_{\textnormal{sup}}$. Therefore, let us suppose that after exiting $D$, we end up in $D_{\textnormal{inf}}$. By leveraging the results of Proposition \ref{exitfinitesingle}, we know that we exit the set $D_{\textnormal{inf}}$ in finite time. Given the way the AoI evolves, it is evident that we can only enter $D$ when we exit $D_{\textnormal{inf}}$. Accordingly, we can conclude that we reach $D$ from $D_{\textnormal{inf}}$ in finite time. To show that this is also the case from $D_{\textnormal{sup}}$, we proceed in a different fashion. Specifically, we consider that $x\in D_{\textnormal{sup}}$ and we first aim to show that we exit $D_{\textnormal{sup}}$ in finite time. In other words, 
\begin{equation}
 \mathbb{E}_{x}[\tau_{D_{\textnormal{sup}}}]<\infty, \quad \forall x\in D_{\textnormal{sup}}. 
\end{equation}
To prove this, we follow the same procedure of the proof of Proposition \ref{exitfinitesingle}. Particularly, let us consider the following twice differentiable function
\begin{equation}
W(x)=x, \quad \forall x\in D_{\textnormal{sup}}.
\end{equation}
We apply the SHS extended generator previously reported in eq. (\ref{shsextendedgenerator}) on this function $W(x)$. We end up with
\begin{equation}
LW(x)=1+\sum_{l=1}^{L}[c-x][\sum_{j=0}^{J_{l}}a^{(l)}_jx^j].
\end{equation}
As discussed in the proof of Lagrange stability, the highest order monomial has a negative coefficient. Accordingly, we have that as $x$ grows, there exists a certain boundary beyond which we get $LW(x)<0$. Therefore, there exists a certain constant $U$ such that if $x>U$, we get $LW(x)<-\epsilon$ for a certain $\epsilon>0$. To that end, let us consider the set $D_{\textnormal{int}}=[\beta,U]\subset D_{\textnormal{sup}}$, and let $\tau_{D_{\textnormal{int}}}$ denote the exit time of $D_{\textnormal{int}}$. If $x\in D_{\textnormal{int}}$, and given the deterministic evolution of the AoI, we have
\begin{equation}
\tau_{D_{\textnormal{int}}}<U-\beta<\infty. 
\end{equation}
In fact, either a transition occurs that reduces the AoI to the value $c$, and hence takes us back to the set $D_{\textnormal{inf}}$, or the AoI keeps growing until we enter the set $D_{\textnormal{sup}}\setminus D_{\textnormal{int}}$ after at most $U-\beta$. Consequently, we have $\tau_{D_{\textnormal{int}}}<\infty$. Recall that if we reach the set $D_{\textnormal{inf}}$, we are guaranteed to enter $D$ in finite time. Therefore, given the strong Markov property, it is sufficient to examine what happens to the process if we start with $x\in D_{\textnormal{sup}}\setminus D_{\textnormal{int}}$. To that end, and as previously explained, we have $W(x)\geq0$ and $LW(x)<-\epsilon$ for a certain $\epsilon>0$ for any $x\in D_{\textnormal{sup}}\setminus D_{\textnormal{int}}$. Therefore,  $W(x)$ is a Lyapunov function. Accordingly, we can proceed in the same way as the proof of Proposition \ref{exitfinitesingle} to show that $\tau_{D_{\textnormal{sup}}\setminus D_{\textnormal{int}}}<\infty$. Given the dynamics of the system, exiting $D_{\textnormal{sup}}\setminus D_{\textnormal{int}}$ takes us back to $D_{\textnormal{inf}}$. By leveraging the strong Markov property, and knowing that from $D_{\textnormal{inf}}$ we enter $D$ in finite time, we can conclude that we are guaranteed to return to $D$ in finite time if we start from $D_{\textnormal{sup}}\setminus D_{\textnormal{int}}$. All in all, for any non-empty open set $D$ with compact closure $\overline{D}$, if we start from any point in $D$, we are guaranteed to return to $D$ in finite time. This concludes our proof. 

\section{Proof of Proposition \ref{exittimemultiple}}
\label{proofexittimemultiple}
For any $(x,q)\in D\times\mathbb{Q}$, let us consider the following function for 
\begin{align}
W^{\boldsymbol{m}}(\boldsymbol{x},q)&=k-\underbrace{(x_1+1)^{m_1}(x_2+1)^{m_2}\ldots(x_{2n}+1)^{m_{2n}}}_{A},
\end{align}
where the constants $k$ and $m_i$ for $i=1,\ldots,2n$ are to be specified. We apply the SHS extended generator previously reported in eq. (\ref{shsextendedgenerator}) on this function $W^{\boldsymbol{m}}(\boldsymbol{x},q)$. By considering $q=0$, we end up with
\begin{equation}
LW^{\boldsymbol{m}}(\boldsymbol{x},0)=-\sum_{i=1}^{n}m_i(x_1+1)^{m_1}\ldots(x_i+1)^{m_i-1}\ldots(x_{2n}+1)^{m_{2n}}.
\end{equation}
Given that $D$ has a compact closure, we can deduce that there exists $U\geq1$ such that
\begin{equation}
1\leq x_i+1\leq U, \quad i=1,\ldots,2n.
\label{inequalitiesonx}
\end{equation}
With that in mind, we can conclude that if $m_i>0$ for $i=1,\ldots,n$, then
\begin{equation}
LW^{\boldsymbol{m}}(\boldsymbol{x},0)\leq-\sum_{i=1}^{n}m_i<0.
\end{equation}
Next, let us examine the SHS extended generator for any state $k\neq0$. Before doing so, and for ease of notation, let us consider
\begin{equation}
A_i=(x_1+1)^{m_1}\ldots(x_i+1)^{m_i-1}\ldots(x_{2n}+1)^{m_{2n}}, \quad i=1,\ldots,2n.
\end{equation}
With this in mind, we get for $k=1,\ldots,n$
\begin{equation}
LW^{\boldsymbol{m}}(\boldsymbol{x},k)=-\sum_{i=1}^{n}m_iA_i-m_{n+k}A_{n+k}+H_k[k-\overline{A}_k-k+A],
\end{equation}
where
\begin{align}
\overline{A}_k=&(x_1+1)^{m_1}\ldots(x_{n+k}+1)^{m_{k}}\ldots(x_{n+k-1}+1)^{m_{n+k-1}}\nonumber\\&(x_{n+k+1}+1)^{m_{n+k+1}}\ldots(x_{2n}+1)^{m_{2n}}.
\end{align}
Given that $\overline{A}_k\geq0$, we can upper bound $LW^{\boldsymbol{m}}(\boldsymbol{x},k)$ as follows
\begin{align}
LW^{\boldsymbol{m}}(\boldsymbol{x},k)&\leq -\sum_{i=1}^{n}m_iA_i-m_{n+k}A_{n+k}+H_kA\nonumber\\&=-\sum_{i=1}^{n}m_i\frac{A}{x_i+1}-m_{n+k}\frac{A}{x_{n+k}+1}+H_kA \nonumber\\&\leq A[H_k-\frac{1}{U}(\sum_{i=1}^{n}m_i+m_{n+k})].
\end{align}
Therefore, it sufficient to choose $\sum_{i=1}^{n}m_i>UH_k$ for the above RHS to be strictly negative for any $k\neq0$. In this case, there exists a sufficiently small $\epsilon>0$ such that
\begin{equation}
LW^{\boldsymbol{m}}(\boldsymbol{x},k)\leq-\epsilon,\quad k=0,1,\ldots,n.
\end{equation}
Lastly, given eq. (\ref{inequalitiesonx}), we can easily choose a constant $k>0$ such that $W^{\boldsymbol{m}}(\boldsymbol{x},q)\geq0$ for any $\boldsymbol{x}\in D$. All in all, we can conclude that $W^{\boldsymbol{m}}(\boldsymbol{x},q)$ is a Lyapunov function. Given that Dynkin's formula is applicable (see \cite[Theorem~1]{HESPANHA20051353}), then we can leverage it along with Fatou’s lemma
as done in the proof reported in Appendix \ref{proofexitfinemnawalwahad} to obtain to obtain
\begin{equation}
\mathbb{E}_{\boldsymbol{x},q}[\tau_D]\leq \frac{W^{\boldsymbol{m}}(\boldsymbol{x},q)}{\epsilon}<\infty.
\end{equation}
\section{Proof of Proposition \ref{qtisreccureent}}
\label{proofqtisreccureent}
As a first step, we can notice from Fig. \ref{generalcsmaxxx} that the chain corresponding to $q(t)$ is irreducible. On top of that, we note that the number of states is finite as it is equal to $n+1$, where $n$ is the number of wireless nodes in the network. Therefore, we can conclude that all states are recurrent (see  \cite[Proposition 2.3]{Cunha13111}). Additionally, let us define $\tau_{\overline{q}\overline{q}}$ as the time for $q(t)$ to return to state $\overline{q}$ given that it started in state $\overline{q}$. By leveraging \cite[Proposition 2.4]{Cunha13111}, we know that it is enough to prove that $\mathbb{E}[\tau_{\overline{q}\overline{q}}]<\infty$ to conclude that all states are positive recurrent and that the expected time to go from any state $q$ to another state $q'$, which we denote by $\mathbb{E}[\tau_{qq'}]$, is also finite. To that end, let us consider that $\overline{q}=\{0\}$, and let us define the following exit time
\begin{equation}
\tau_{0}:=\inf\{t\geq0: q(t)\neq0\},
\end{equation}
given that $q(0)=0$. Now, given the dynamics of the age processes reported in Section \ref{lagrangeeee}, we know that the age at the monitor of each link $i$ increases linearly with time. On top of that, we recall that we have $x_i(t)\neq0$ for $i=1,\ldots,n$. 
Note that a transition happens when any of the $n$ exponential clocks with rates $a_ix_i(t)$ ticks. To that end, we have that a tick happens with a rate larger than $\underset{i}{\text{min }}a_ix_i(0)>0$. Given that the clocks are exponential, we can conclude that this happens with an average time smaller than $\frac{1}{\underset{i}{\text{min }}a_ix_i(0)}<\infty$. Accordingly, we have
\begin{equation}
\mathbb{E}_{\boldsymbol{x}}[\tau_{0}]\leq \frac{1}{\underset{i}{\text{min }}a_ix_i(0)}<\infty.
\end{equation}
Now, let us suppose that the transition takes us to any of the states $k\neq0$. By construction, we go back to state $0$ in an exponentially distributed time with rate $H_k$. Hence, we conclude that for any initial continuous state $\boldsymbol{x}(0)$
\begin{equation}
\mathbb{E}_{\boldsymbol{x}}[\tau_{00}]\leq \frac{1}{\underset{i}{\text{min }}a_ix_i(0)}+\frac{1}{\underset{i}{\text{min }}H_i}<\infty.
\end{equation}
Consequently, the stochastic process $\{q(t):t\geq0\}$ is positive recurrent.
\section{Proof of Theorem \ref{ahamtheoremblpositive}}
\label{proofahamtheoremblpositive}
Let us start our analysis by looking at the transitions out of state $0$. Specifically, we have that the rate of these transitions is equal to $a_ix_i(t)$ for $i=1,\ldots,n$. Therefore, starting from any time instant $t_0$, let $G_i$ be an exponential clock of rate $a_ix_i(t)$ and let us investigate the probability that a tick takes place after a certain time $\epsilon>0$ elapsed from $t_0$. Using the notion of survivor function and hazard rate (see \cite[Chapter~2]{a._199322}), we obtain
\begin{equation}
\Pr(G_i>\epsilon)=\exp[-\int_{t_0}^{t_0+\epsilon}a_ix_i(t)dt]\overset{\Delta}{=} p_i(\epsilon)>0,\quad i=1,\ldots,n,
\label{azgharmnwehde}
\end{equation}
where the strict positivity is a consequence of Lemma \ref{differentthan0}. Given the above expression, we can conclude that for $\epsilon_1<\epsilon_2$
\begin{align}
\Pr(\epsilon_1<G_i<\epsilon_2)&=\exp[-\int_{t_0}^{t_0+\epsilon_1}a_ix_i(t)dt]\nonumber\\&-\exp[-\int_{t_0}^{t_0+\epsilon_2}a_ix_i(t)dt]>0,\quad i=1,\ldots,n,
\label{bentnen}
\end{align}
where the strict positivity comes from the fact that the age of each link grows linearly with time in state $0$. We recall that our goal is to show that every set $U=D\times\{q\}\in D_{\textnormal{definition}}\times\mathbb{Q}$ where $D$ is a non-empty open set with compact closure is positive recurrent. First, we note that, by definition, any open set $D$ can be considered as the union of open balls. Therefore, it is sufficient to prove the positive recurrence of any set of the form $U=B(\overline{\boldsymbol{x}},\epsilon)\times\{q\}$ where $B(\overline{\boldsymbol{x}},\epsilon)\subset D_{\textnormal{definition}}$ is a ball of center $\overline{\boldsymbol{x}}$ and radius $\epsilon>0$. With this in mind, we recall that the vector space to which $\boldsymbol{x}(t)$ belongs is finite dimensional. Therefore, all norms are equivalent and, accordingly, let us consider the $\infty$-norm $||\boldsymbol{x}(t)||_{\infty}=\underset{i}{\text{max}}\:x_i(t)$. Using this norm, the ball $B(\overline{\boldsymbol{x}},\epsilon)$ is a square shaped open set. Next, given the particularity of the domain $D_{\textnormal{definition}}$, we detail below the form that $B(\overline{\boldsymbol{x}},\epsilon)$ can have. For example, in state $q=\{0\}$, $B(\overline{\boldsymbol{x}},\epsilon)$ can be rewritten as the set of $\boldsymbol{x}$ such that 
\begin{align}
a_1<&x_1<b_1,\nonumber\\
a_2<&x_2<b_2, \nonumber\\
&\vdots \nonumber\\
a_n<&x_n<b_n, \nonumber\\
&\hspace{-25pt}x_i=0, \quad i=n+1,\ldots,2n,
\end{align}
where $b_i-a_i=2\epsilon>0$. Note that the above set is open in $\mathbb{R}^n$. Similarly, in state $q=\{k\}$, we can equivalently consider the set 
\begin{align}
a_1<&x_1<b_1,\nonumber\\
a_2<&x_2<b_2, \nonumber\\
&\vdots \nonumber\\
a_n<&x_n<b_n, \nonumber\\
a_{n+k}<&x_{n+k}<b_{n+k}, \nonumber\\
&\hspace{-25pt}x_i=0, \quad i=n+1,\ldots,2n,
\end{align}
such that $b_i-a_i=2\epsilon>0$ and $x_{n+k}<x_i$ for $i=1,\ldots,n$. Note that this set is open in $\mathbb{R}^{n+1}$. Let us now investigate the recurrence time of the set $U=B(\overline{\boldsymbol{x}},\epsilon)\times\{k\}$, denoted by $\tau_{UU}$. To do so, let us suppose that we start from a point $\boldsymbol{x}\in B(\overline{\boldsymbol{x}},\epsilon)$ and $q(0)=k$. Without loss of generality, we suppose that $a_1\geq a_2 \geq\ldots\geq a_n>a_{n+k}>0$. Note that $a_{n+k}$ is always strictly smaller than $a_i$ for $i=1,\ldots,n$ since $x_{n+k}<x_i$. The recurrence time $\tau_{UU}$ can be decomposed into two components: the exit time of $U$ and the reentry time in $U$. By Proposition \ref{exittimemultiple}, we have that the exit time $\tau_{B(\overline{\boldsymbol{x}},\epsilon)}$ (and consequently $U$) has a finite expectation. Given that the exit time of $U$ is a stopping time, we can leverage the strong Markov property and suppose that we have started outside of $U$. Then, we can focus on proving that the expectation of the reentry time in $U$ is finite to prove the positive recurrence with respect to $U$. Consequently, let us consider that $\boldsymbol{x}(0)\notin B(\overline{\boldsymbol{x}},\epsilon)$ and $q(0)\in\mathbb{Q}$. Additionally, let us define the following family of stopping times
\begin{align}
\sigma^1_0&=\inf\{t\geq0: q(t)=0\},\nonumber\\
\sigma^1_1&=\inf\{t\geq\sigma^1_0: q(t)=1\},\nonumber\\ \sigma^1_2&=\inf\{t\geq\sigma^1_1: q(t)=0\},\nonumber\\ \sigma^1_3&=\inf\{t\geq\sigma^1_2: q(t)=2\},\nonumber\\ &\vdots\nonumber\\ \sigma^1_{2n-1}&=\inf\{t\geq\sigma^1_{2n-2}: q(t)=n\},\nonumber\\ \sigma^1_{2n}&=\inf\{t\geq\sigma^1_{2n-1}: q(t)=0\},\nonumber\\ 
\sigma^1_{2n+1}&=\inf\{t\geq\sigma^1_{2n}: q(t)=k\},\nonumber\\\sigma^1_{2n+2}&=\inf\{t\geq\sigma^1_{2n+1}: q(t)=0\},\nonumber\\  
\sigma^2_{1}&=\inf\{t\geq\sigma^1_{2n+2}: q(t)=1\},\nonumber\\ 
\sigma^2_{2}&=\inf\{t\geq\sigma^2_{1}: q(t)=0\},\nonumber\\
 &\vdots\nonumber\\ \sigma^N_{2i-1}&=\inf\{t\geq\sigma^N_{2i-2}: q(t)=i\},\nonumber\\ \sigma^N_{2i}&=\inf\{t\geq\sigma^N_{2i-1}: q(t)=0\},\nonumber\\  &\vdots
\end{align}
Basically, we track the following trajectory of $q(t)$
\begin{align}
&q(0)\rightarrow0\rightarrow1\rightarrow0\rightarrow2\rightarrow\ldots\rightarrow k \rightarrow0\rightarrow\ldots\nonumber\\&\rightarrow n\rightarrow 0\rightarrow k\rightarrow0 \rightarrow1\rightarrow\ldots.
\end{align}
Using the above trajectory, we will establish the positive recurrence of the process $\{\big(\boldsymbol{x}(t),q(t)\big): t\geq0\}$ . To do so, we first recall that $\{q(t):t\geq0\}$ is positive recurrent. With this in mind, and given the strong Markov property of the stochastic process $\{\big(\boldsymbol{x}(t),q(t)\big): t\geq0\}$, we have
\begin{equation}
\mathbb{E}_{\boldsymbol{x},q}[\sigma^1_{0}]<\infty,
\end{equation}
\begin{align}
\mathbb{E}_{\boldsymbol{x}(\sigma^N_{2i-2})}[\sigma^N_{2i-1}-\sigma^N_{2i-2}]<\infty,&\quad \forall \boldsymbol{x}(\sigma^N_{2i-2})\in[0,\infty)^{2n},\nonumber\\&\forall i\in\{1,\ldots,n+1\}, \forall N\in\mathbb{N}^{*}.
\end{align}
Essentially, the expected time between these stopping times is finite, whatever the value of $\boldsymbol{x}$ at the end of the previous stopping time is. Next, we can notice that $\sigma^N_{2i}-\sigma^N_{2i-1}$ is exponentially distributed with rate $H_i$. Accordingly, we conclude that
\begin{align}
\mathbb{E}_{\boldsymbol{x}(\sigma^N_{2i-1})}[\sigma^N_{2i}-\sigma^N_{2i-1}]=\frac{1}{H_i}<\infty,&\quad \forall \boldsymbol{x}(\sigma^N_{2i-1})\in[0,\infty)^{2n},\nonumber\\&\forall i\in\{1,\ldots,n\},\forall N\in\mathbb{N}^{*},
\end{align}
\begin{equation}
\Pr[\epsilon_1<\sigma^N_{2i}-\sigma^N_{2i-1}<\epsilon_2]=\exp[-H_i\epsilon_1]-\exp[-H_i\epsilon_2]>0,
\label{eqq22}
\end{equation}
\begin{align}
\mathbb{E}_{\boldsymbol{x}(\sigma^N_{2n+1})}[\sigma^N_{2n+2}-\sigma^N_{2n+1}]=\frac{1}{H_k}<\infty,\quad &\forall \boldsymbol{x}(\sigma^N_{2n+1})\in[0,\infty)^{2n},\nonumber\\&\forall N\in\mathbb{N}^{*},
\end{align}
\begin{equation}
\Pr[\epsilon_1<\sigma^N_{2n+2}-\sigma^N_{2n+1}<\epsilon_2]=\exp[-H_k\epsilon_1]-\exp[-H_k\epsilon_2]>0,
\label{eqq24}
\end{equation}
for $\epsilon_1<\epsilon_2$. All in all, we have shown that the expectation between each pair of stopping times is finite. Next, we aim to show that the exact same order of the trajectory can be followed within a specified elapsed time. To that end, let us define for every $\epsilon>0$ the following events 
\begin{align}
A^N_{2i-1}=&\{\sigma^N_{2i-1}-\sigma^N_{2i-2}<\frac{\epsilon}{2n}\bigcap G_j>\frac{\epsilon}{2n}\textnormal{ for } j\neq i\},\nonumber\\ &\forall i\in\{1,\ldots,n\},\forall N\in\mathbb{N}^{*},\nonumber\\
A^N_{2n+1}=&\{\sigma^N_{2n+1}-\sigma^N_{2n}<\frac{\epsilon}{2n}\bigcap G_j>\frac{\epsilon}{2n}\textnormal{ for } j\neq k\},\nonumber\\& \forall N\in\mathbb{N}^{*},\nonumber\\
A^N_{2i}=&\{a_i-a_{i+1}<\sigma^N_{2i}-\sigma^N_{2i-1}<a_i-a_{i+1}+\frac{\epsilon}{2n}\},\nonumber\\& \forall i\in\{1,\ldots,n-1\},\forall N\in\mathbb{N}^{*},\nonumber\\
A^N_{2n}=&\{a_n-a_{n+k}<\sigma^N_{2n}-\sigma^N_{2n-1}<a_n-a_{n+k}+\frac{\epsilon}{2n}\},\nonumber\\&\forall N\in\mathbb{N}^{*},\nonumber\\
A^N_{2n+2}=&\{\sigma^N_{2n+2}-\sigma^N_{2n+1}>a_n\},\quad \forall N\in\mathbb{N}^{*}.
\end{align}
In other words, the event $A^N_{2i-1}$ for $i=1,\ldots,n$ is true when the only transition that takes place between $\sigma^N_{2i-1}-\sigma^N_{2i-2}$ is the transition from state $0$ to state $i$. Additionally, this transition happens in a time less than $\frac{\epsilon}{2n}$. Given eq. (\ref{azgharmnwehde}), and by noting the independence between the different exponential clocks in state $q=0$, we can conclude that for $i=1,\ldots,n$
\begin{equation}
{\Pr}_{\boldsymbol{x}(\sigma^N_{2i-2})}(A^N_{2i-1})=(1-p_i(\frac{\epsilon}{2n}))\prod_{j\neq i}p_j(\frac{\epsilon}{2n})\overset{\Delta}{=}\eta_i(\frac{\epsilon}{2n})>0,
\end{equation}
and
\begin{equation}
{\Pr}_{\boldsymbol{x}(\sigma^N_{2n})}(A^N_{2n+1})=(1-p_k(\frac{\epsilon}{2n}))\prod_{j\neq k}p_j(\frac{\epsilon}{2n})\overset{\Delta}{=}\eta_k(\frac{\epsilon}{2n})>0.
\end{equation}
As for the event $A^N_{2i}$ for $i=1,\ldots,n-1$, it is true when the transmission time is in the interval $]a_i-a_{i+1},a_i-a_{i+1}+\frac{\epsilon}{2n}[$. Given eq. (\ref{eqq22}) and (\ref{eqq24}), we have for all $N\in\mathbb{N}^*$ 
\begin{align}
&{\Pr}_{\boldsymbol{x}(\sigma^N_{2i-1})}(A^N_{2i})=\exp[-H_i(a_i-a_{i+1})]\nonumber\\&-\exp[-H_i(a_i-a_{i+1}+\frac{\epsilon}{2n})]\overset{\Delta}{=} \delta_i(\frac{\epsilon}{2n})>0,\: i=1,\ldots,n-1,\\
&{\Pr}_{\boldsymbol{x}(\sigma^N_{2n-1})}(A^N_{2n})=\exp[-H_i(a_n-a_{n+k})]\nonumber\\&-\exp[-H_i(a_n-a_{n+k}+\frac{\epsilon}{2n})]\overset{\Delta}{=} \delta_{n}(\frac{\epsilon}{2n})>0,\\ 
&{\Pr}_{\boldsymbol{x}(\sigma^N_{2n+1})}(A^N_{2n+2})=\exp[-H_ka_{n+k}]\overset{\Delta}{=} \delta_{n_k}>0. 
\end{align}
We can see that if all the mentioned events are true for a particular cycle of transitions $N$, then we are guaranteed to return to the set $U$ at a time $\tau$ such that $\sigma^N_{2n+1}\leq\tau\leq \sigma^N_{2n+2}$. In fact, as it has been discussed in Section \ref{recurrenceeee}, upon a transition back to state $0$ from a state $i$, the age of link $i$ becomes equal to the time spent in state $i$ (i.e., equal to $x_{i+n}$ while $x_{i+n}$ becomes equal to $0$). To that end, let us denote by 
\begin{equation}
N^*=\inf\{N\in\mathbb{N}^*:A^N_{2i} \textnormal{ and }A^N_{2i-1} \textnormal{ are true for }i=1,\ldots,n+1\},
\end{equation}
the first transition cycle such that the events $A^N_{2i}$ and $A^N_{2i-1}$ are true for $i=1,\ldots,n+1$. Consequently, we have
\begin{equation}
\tau\leq\sigma^{N^*}_{2n+2}. 
\end{equation} 
Therefore, it is sufficient to show that $\mathbb{E}_{\boldsymbol{x},q}[\sigma^{N^*}_{2n+2}]<\infty$ to conclude the positive recurrence of the stochastic process $\{\big(\boldsymbol{x}(t),q(t)\big): t\geq0\}$. To that end, let $\displaystyle\rho=\min_{i}\{\eta_i,\delta_i\}>0$. Given the strong Markov property of $\{\big(\boldsymbol{x}(t),q(t)\big): t\geq0\}$, we end up with 
\begin{equation}
{\Pr}_{\boldsymbol{x}(0)}[N^*=\infty]\leq\lim_{N\to\infty}[1-\rho^{2n+2}]^N=0.
\end{equation}
In other words, we are guaranteed that at some point, all the events will come true in a single cycle $N$. Next, given that the expected time between each stopping time is finite, we can upper bound them all by a constant $M>0$. Accordingly, the expected time elapsed at every cycle of transitions $N\geq2$ is upper bounded by $(2n+2)M$. For $N=1$, the expected time elapsed is upper bounded by $(2n+3)M$. With this in mind, and given the strong Markov property of the stochastic process $\{\big(\boldsymbol{x}(t),q(t)\big): t\geq0\}$, we can conclude that
\begin{align}
\mathbb{E}_{\boldsymbol{x},q}[\sigma^{N^*}_{2n+2}]&=\sum_{N=1}^{\infty}{\Pr}_{\boldsymbol{x}(0)}[N^*=N]\mathbb{E}_{\boldsymbol{x},q}[\sigma^{N^*}_{2n+2}|N^{*}=N]\nonumber\\&\leq \sum_{N=1}^{\infty}[1-\rho^{2n+2}]^N(N\times (2n+3)M)< \infty,
\label{equationno2tano2ta}
\end{align}
for all $\boldsymbol{x}\notin B(\overline{\boldsymbol{x}},\epsilon)$ and $q\in\mathbb{Q}$. Combining this with the fact that the expected exit time of $U$ is finite, we can conclude that the expectation of the recurrence time $\tau_{UU}$ is finite. This concludes our proof.

\section{Proof of Theorem \ref{stabilitymultipleee}}
\label{ekherproof}
To proceed with our proof, let us consider an arbitrary time instant $t$ ad let us suppose that the process $\{\big(\boldsymbol{x}(t),q(t)\big): t\geq0\}$ starts from the initial state $(\boldsymbol{x}(0),q(0))$. Let us also consider the open ball $D=B(\boldsymbol{x}(0),\epsilon)\subset D_{\textnormal{definition}}$ centered around $\boldsymbol{x}(0)$ of radius $\epsilon>0$. We suppose that for each point $\boldsymbol{x}\in B(\boldsymbol{x}(0),\epsilon)$, we have $x_i\leq U$ for $i=1,\ldots,2n$, where $U$ is a finite positive number. We recall from Theorem \ref{ahamtheoremblpositive} that the stochastic process $\boldsymbol{x}(t)$ is positive recurrent with respect to $D$. To that end, let us define $N^*$ such that
\begin{equation}
\tau^{N^*}_{DD}<t\leq\tau^{N^*+1}_{DD},
\end{equation}
where $\tau^{N}_{DD}$ is the $N$-th recurrence time of the set $D$. More precisely, $\tau^{1}_{DD}$ is the time after which the process returns to $D$ after leaving it for the first time. Similarly, $\tau^{2}_{DD}$ is the time the process returns to $D$ after it leaves it for the \emph{second} time. Now, given that $D$ is bounded and that $x_i(t)$ increases at most linearly with time, we have
\begin{equation}
x_i(t)\leq U+\underbrace{\tau^{N^*+1}_{DD}-\tau^{N^*}_{DD}}_{C_{N^*}},\quad i=1,\ldots,2n.
\end{equation}
In essence, $C_{N^*}$ is the time elapsed between the two consecutive recurrence times. Accordingly, if we are interested in the moment of $\boldsymbol{x}(t)$ associated with $\boldsymbol{m}$, previously denoted as $\mu^{\boldsymbol{m}}(t)=\mathbb{E}[\boldsymbol{x}^{\boldsymbol{m}}(t)]$, then we have
\begin{equation}
\mu^{\boldsymbol{m}}(t)\leq \mathbb{E}[(U+C_{N^*})^m]=\mathbb{E}[\sum_{i=0}^{m}{m\choose i}C_{N^*}^iU^{m-i}],
\end{equation}
where $m=\sum_{i=1}^{2n}m_i$. Next, we recall the results of eq. (\ref{illustrumomentcomparisonconvex}) where Jensen's inequality was leveraged. Using these results, we can conclude that if $\mathbb{E}[C_{N^*}^m]$ is finite, then the same can be said for $\mathbb{E}[C_{N^*}^p]$ for $p<m$. Therefore, to prove that $\mathbb{E}[(U+C_{N^*})^m]$ is finite, and given that $U$ is a constant, it is sufficient to show that $\mathbb{E}[C_{N^*}^m]<\infty$.

Now that we know our goal, we can proceed with our proof. To that end, we recall our proof of Theorem \ref{ahamtheoremblpositive} where we have considered the recurrence time of any non-empty open set of $D_{\textnormal{definition}}$ with compact closure. By following the same procedure, we can consider a specific trajectory of $q(t)$ to reach the set $B(\boldsymbol{x}(0),\epsilon)$. By doing so, and by leveraging the strong Markov property, we can obtain similar results to eq. (\ref{equationno2tano2ta}). In essence, to show that the moment of order $m$ of $C_{N^*}$ is finite, it suffices to show that the moment of order $m$ of the transition time between any two discrete states $i\neq j$ is finite. To do so, let $T^{\boldsymbol{x}}_{ij}$ denote the first time that the discrete process $q(t)$ visits state $j$ if it started in $i$, given the initial state $\boldsymbol{x}$. Using the results reported in \cite{10.1214/ECP.v16-1632}, we can conclude that if $\mathbb{E}_{\boldsymbol{x}}[\tau^m_{00}]<\infty$ then $\mathbb{E}_{\boldsymbol{x}}[T^{\boldsymbol{x}}_{ij}]<\infty$, where $\tau^m_{00}$ is the moment of order $m$ of the recurrence time of the discrete state $0$. With that in mind, we note that the recurrence time can be divided into two components: the exit time of state $0$ and the return time to state $0$. Therefore, we have
\begin{equation}
\mathbb{E}_{\boldsymbol{x}}[\tau^m_{00}]=\mathbb{E}_{\boldsymbol{x}}[(\tau_{\textnormal{exit}}+\tau_{\textnormal{return}})^m]=\mathbb{E}_{\boldsymbol{x}}[\sum_{i=0}^{m}{m\choose i}\tau_{\textnormal{exit}}^i\tau_{\textnormal{return}}^{m-i}].
\end{equation}
Let us investigate the exit time of state $0$. Following the same analysis found in the proof of Proposition \ref{qtisreccureent}, we can conclude that
\begin{equation}
\mathbb{E}_{\boldsymbol{x}}[\tau_{\textnormal{exit}}^m]\leq \mathbb{E}[Y^m],
\end{equation}
where $Y$ is an exponential random variable of rate $\underset{i}{\text{min }}a_ix_i(0)>0$. Given that $\mathbb{E}[Y^m]$ is finite for any $m>0$, we can conclude that the same goes for $\mathbb{E}_{\boldsymbol{x}}[\tau_{\textnormal{exit}}^m]$. Lastly, we focus on the return time to state $0$. To that end, we note that when we exit the state $0$, we end up in any of the remaining states $k\neq0$. Given that the time spent in state $k$ is exponentially distributed with mean $\frac{1}{H_k}$, and given the independence between $\tau_{\textnormal{exit}}$ and $\tau_{\textnormal{return}}$, we can conclude that 
\begin{equation}
\mathbb{E}_{\boldsymbol{x}}[\tau^m_{00}]<\infty, \quad m>0. 
\end{equation}
With this in mind, we can conclude that $\mathbb{E}[C_{N^*}^m]<\infty$. Accordingly, the moment of $\boldsymbol{x}(t)$ associated with $\boldsymbol{m}$ such that $m_i\in\mathbb{N}$ for $i=1,\ldots,2n$ and $\sum_{i=1}^{2n}m_i=m$ is finite for any arbitrary time instant $t$. 

\section{Proof of Proposition \ref{finalconvergence}}
\label{appendixfinalconvergence}
To prove the proposition, we need to showcase several key properties of  
the SCA function $\overline{\Upsilon}(\boldsymbol{x},\boldsymbol{y})$ in (\ref{scafunction}) and leveraging them to demonstrate the intended results. The characteristics are summarized in the following:
\begin{itemize}
\item $\overline{\Delta}(\boldsymbol{x})\leq\overline{\Upsilon}(\boldsymbol{x},\boldsymbol{y}),\quad \forall \boldsymbol{x},\boldsymbol{y}\in\mathcal{X}$
\item $
\lim_{\boldsymbol{x}\to \boldsymbol{y}}\nabla_{\vec{\boldsymbol{d}}}\overline{\Upsilon}(\boldsymbol{x},\boldsymbol{y})=\nabla_{\vec{\boldsymbol{d}}}\overline{\Delta}(\boldsymbol{y}), \quad \forall \vec{\boldsymbol{d}},\boldsymbol{y}:\boldsymbol{y},\vec{\boldsymbol{d}}+\boldsymbol{y}\in\mathcal{X}$
\item $\overline{\Upsilon}(\boldsymbol{x},\boldsymbol{y})$ is continuous $\forall(\boldsymbol{x},\boldsymbol{y})\in\mathcal{X}\times\mathcal{X}$
\item $\overline{\Upsilon}(\boldsymbol{y},\boldsymbol{y})=\overline{\Delta}(\boldsymbol{y}),\quad \forall \boldsymbol{y}\in\mathcal{X}$
\end{itemize}
If the above properties are verified, and knowing that $\mathcal{X}$ is a compact set, we can assert that the assumptions in \cite[Assumption 1]{doi:10.1137/120891009} are true. Accordingly, we can use \cite[Theorem 1]{doi:10.1137/120891009} to show that every limit point of the iterates generated by the algorithm in (\ref{scaprocedure}) is a stationary point of the problem in (\ref{Nusersopt}).\\
Let us start by examining the first property. To that end, we recall the objective function below
\begin{equation}
\overline{\Delta}(\boldsymbol{x})=c\boldsymbol{d}^T\boldsymbol{E}^{-1}(\boldsymbol{x})\boldsymbol{b}_{\infty|m}(\boldsymbol{x}).
\end{equation}
By construction, the closed linear system being examined is fully determined, and hence $\boldsymbol{E}^{-1}(\boldsymbol{x})$ exists. Therefore, we have that
\begin{equation}
\big(\boldsymbol{E}^{-1}\big)_{ij}=\frac{(-1)^{i+j}M_{ji}}{\det(\boldsymbol{E})},
\end{equation}
where $M_{ji}$ is the $(j,i)$ minor of $\boldsymbol{E}$. Given that the transition rates are polynomial functions, we can conclude that the entries of $\big(\boldsymbol{E}\big)_{ij}$ are monomials of the components of $\boldsymbol{x}$. With that in mind, we can conclude that the entries $\big(\boldsymbol{E}^{-1}\big)_{ij}$ are all rational functions of $\boldsymbol{x}$ with the denominator being $\det(\boldsymbol{E})$. Therefore, one can deduce that the overall age function $\overline{\Delta}(\boldsymbol{x})$ can be written as a rational function of $\boldsymbol{x}$ with non-zero denominator for any $\boldsymbol{x}\in\mathcal{X}$. Consequently, we can affirm that $\overline{\Delta}(\boldsymbol{x})$ is a continuous and differentiable function on the set $\mathcal{X}$. The same argument can be made for the gradient function $\nabla\overline{\Delta}(\boldsymbol{x})$ and, accordingly, the entries of the Hessian matrix $\nabla^{2}\overline{\Delta}(\boldsymbol{x})$. Due to the continuity of the Hessian matrix, and given that $\mathcal{X}$ is compact, we know that there exist a constant $L>0$ such that $||\nabla^{2}\overline{\Delta}(\boldsymbol{x})||\leq L$ for $\boldsymbol{x}\in\mathcal{X}$. By using the mean value theorem on the function $\nabla\overline{\Delta}(\boldsymbol{x})$, and by noting the bound on $||\nabla^{2}\overline{\Delta}(\boldsymbol{x})||$, we can show that $\nabla\overline{\Delta}(\boldsymbol{x})$ is a Lipschitz function. More specifically, there exists a constant $L>0$ such that
\begin{equation}
||\nabla\overline{\Delta}(\boldsymbol{x})-\nabla\overline{\Delta}(\boldsymbol{y})||\leq L ||\boldsymbol{x}-\boldsymbol{y}||, \quad \forall \boldsymbol{x},\boldsymbol{y}\in\mathcal{X}.
\label{lipz}
\end{equation}
Using the above results, we can apply the descent Lemma \cite[Proposition A.24]{Bertsekas/99} to show that $\overline{\Delta}(\boldsymbol{x})\leq\overline{\Upsilon}(\boldsymbol{x},\boldsymbol{y}),\:\:\forall\boldsymbol{x},\boldsymbol{y}\in\mathcal{X}$ if $\alpha_n\geq \frac{L}{2}$. \\
Next, we investigate the second property revolving around the directional derivative of the function $\overline{\Upsilon}(\boldsymbol{x},\boldsymbol{y})$ with respect to $\boldsymbol{x}$. More specifically
\begin{equation}
\nabla_{\vec{\boldsymbol{d}}}\overline{\Upsilon}(\boldsymbol{x},\boldsymbol{y})=\lim_{\lambda \to 0} \frac{\overline{\Upsilon}(\boldsymbol{x}+\lambda\vec{\boldsymbol{d}},\boldsymbol{y})-\overline{\Upsilon}(\boldsymbol{x},\boldsymbol{y})}{\lambda}.
\end{equation}
By replacing $\overline{\Upsilon}(\boldsymbol{x},\boldsymbol{y})$ with its value from eq. (\ref{scafunction}), we can show that $\lim_{\boldsymbol{x}\to \boldsymbol{y}}\nabla_{\vec{\boldsymbol{d}}}\overline{\Upsilon}(\boldsymbol{x},\boldsymbol{y})=\langle\nabla\overline{\Delta}(\boldsymbol{y})^T,\vec{\boldsymbol{d}}\rangle=\nabla_{\vec{\boldsymbol{d}}}\overline{\Delta}(\boldsymbol{y})$, where $\langle\cdot,\cdot\rangle$ denotes the dot product. \\
As for the third property, we can easily show that it holds by noting the expression of $\overline{\Upsilon}(\boldsymbol{x},\boldsymbol{y})$ in (\ref{scafunction}) and by taking into account that $\overline{\Delta}(\boldsymbol{x})$ and $\nabla\overline{\Delta}(\boldsymbol{x})$ were shown to be continuous $\forall\boldsymbol{x}\in\mathcal{X}$. Lastly, by simple substitution, we can assert that $\overline{\Upsilon}(\boldsymbol{y},\boldsymbol{y})=\overline{\Delta}(\boldsymbol{y}),\:\forall \boldsymbol{y}\in\mathcal{X}$. Given that all the above properties are verified, we can confirm that the limit point of the sequence $\big\{\overline{\boldsymbol{\eta}}[k]\big\}_{k=1}^{+\infty}$ is a stationary point of the problem in (\ref{Nusersopt11}) if $\alpha_k\geq \frac{L}{2},\:k\in\mathbb{N}$.
\end{document}